\documentclass[twocolumn]{aastex62}
\usepackage{amsmath}

\graphicspath{{./}{figures/}}
\newcommand{\mgfe}[0]{[{\rm Mg/Fe}]} 
\newcommand{\Acc}{A_{\rm cc}}
\newcommand{\AIa}{A_{\rm Ia}} 
\newcommand{\RIa}{R_{\rm Ia}^X}
\newcommand{\aIa}{\alpha_{\rm Ia}}
\newcommand{\acc}{\alpha_{\rm cc}}
\newcommand{\afe}[0]{[\alpha/{\rm Fe}]}
\newcommand{\femg}{[{\rm Fe}/{\rm Mg}]} 
\newcommand{\xmg}{[{\rm X}/{\rm Mg}]} 
\newcommand{\mgh}{[{\rm Mg}/{\rm H}]}
\newcommand{\feh}[0]{[{\rm Fe/H}]} 
\newcommand{\pIa}{p_{\rm Ia}^{\rm X}}
\newcommand{\pcc}{p_{\rm cc}^{\rm X}}
\newcommand{\fcc}{f_{\rm cc}}

\defcitealias{weinberg}{W19}
\defcitealias{andrews}{AWSJ17}
\defcitealias{buder_a}{B18}
\defcitealias{buder_b}{B19}
\defcitealias{tuguldur}{S16}

\submitjournal{ApJ}

\shorttitle{Abundance ratios in GALAH DR2}
\shortauthors{Griffith et al.}

\begin{document}

\title{Abundance ratios in GALAH DR2 and their implications for nucleosynthesis}

\correspondingauthor{Emily Griffith}
\email{griffith.802@osu.edu}

\author{Emily Griffith}
\affil{The Department of Astronomy and Center of Cosmology and AstroParticle Physics, The Ohio State University, Columbus, OH 43210, USA}

\author{Jennifer A. Johnson}
\affil{The Department of Astronomy and Center of Cosmology and AstroParticle Physics, The Ohio State University, Columbus, OH 43210, USA}

\author{David H. Weinberg}
\affil{The Department of Astronomy and Center of Cosmology and AstroParticle Physics, The Ohio State University, Columbus, OH 43210, USA}



\begin{abstract}

Using a sample of 70 924 stars from the second data release of the GALAH optical spectroscopic survey, we construct median sequences of $\xmg$ vs. $\mgh$ for 21 elements, separating the high-$\alpha$/``low-Ia'' and low-$\alpha$/``high-Ia'' stellar populations through cuts in $\mgfe$. Previous work with the near-IR APOGEE survey has shown that such sequences are nearly independent of location in the Galactic disk, implying that they are determined by stellar nucleosynthesis yields with little sensitivity to other chemical evolution aspects. The separation between the two [X/Mg] sequences indicates the relative importance of prompt and delayed enrichment mechanisms, while the sequences' slopes indicate metallicity dependence of the yields. GALAH and APOGEE measurements agree for some of their common elements, but differ in sequence separation or metallicity trends for others. GALAH offers access to nine new elements. We infer that about $75\%$ of solar C comes from core collapse supernovae and $25\%$ from delayed mechanisms. We find core collapse fractions of $60-80\%$ for the Fe-peak elements Sc, Ti, Cu, and Zn, with strong metallicity dependence of the core collapse Cu yield. For the neutron capture elements Y, Ba, and La, we infer large delayed contributions with non-monotonic metallicity dependence. The separation of the [Eu/Mg] sequences implies that at least $\sim30\%$ of Eu enrichment is delayed with respect to star formation. We compare our results to predictions of several supernova and AGB yield models; C, Na, K, Mn, and Ca all show discrepancies with models that could make them useful diagnostics of nucleosynthesis physics.
\end{abstract}


\section{Introduction} \label{sec:intro}

Like a Galactic archaeological record, stellar spectra preserve chemical signatures of the interstellar medium (ISM) at the time of their birth. Predictions of Galactic Chemical Evolution (GCE) models depend on their descriptions of the gas accretion, star formation, metal outflow, and dynamical history of the Galaxy, and equally on the nucleosynthetic yields adopted for element production by different classes of stars. A new generation of giant spectroscopic surveys, most notably \textit{Gaia}-ESO \citep{gilmore2012}, APOGEE\footnote{APOGEE = Apache Point Observatory Galactic Evolution Experiment, originally part of the Sloan Digital Sky Survey III \citep[SDSS-III;][]{eisenstein2011} and continuing under SDSS-IV \citep{blanton2017}} \citep{majewski2017}, and GALAH\footnote{GALAH = GALactic Archaeology with HERMES} \citep{de_silva, martell} are now providing detailed chemical fingerprints for hundreds of thousands of stars in all regions of the Milky Way, a vast increase over earlier high-resolution spectroscopic surveys that targeted $\sim 10^3$ stars primarily in the solar vicinity \citep[e.g.,][]{nissen2010, adibekyan, bensby}. Separating constraints on Galactic history from uncertainties in stellar yields is one of the principal challenges in interpreting these powerful high-dimensional data sets.  

A recent study based on APOGEE data \citep[][hereafter W19]{weinberg} outlines a novel approach to this problem. Examining a sample of 20 485 luminous giant stars spanning Galactocentric radius $R=3-15$ kpc and midplane distance $|Z|=0-2$ kpc, \citetalias{weinberg} find that the median trends of abundance ratios $\xmg$ vs. $\mgh$ are nearly independent of location in the disk, provided one separates the distinct populations of stars with high and low $\femg$ values (see Equation~\ref{eq:boundary} below)\footnote{We follow standard logarithmic, solar-normalized notation for abundances ratios, $[\text{X}/\text{Y}] = \log_{10}(\text{X}/\text{Y}) - \log_{10}(\text{X}/\text{Y})_{\odot}$.}. \citetalias{weinberg} choose Mg rather than Fe as their reference element because Mg is thought to come almost entirely from core collapse supernovae (CCSN), while at solar $\mgfe \approx 0$ the Fe comes from CCSN and Type Ia supernovae (SNIa) in roughly equal measure. The universality of the observed median trends over regions with radically different star formation and enrichment histories implies that these trends are driven by nucleosynthesis physics in ways that are insensitive to other aspects of chemical evolution. \citetalias{weinberg} fit these trends with a simple ``2-process'' model that characterizes the relative production of CCSN and SNIa for each of the elements that they consider: O, Na, Mg, Al, Si, P, S, K, Ca, V, Cr, Mn, Fe, Co, Ni.

In this paper we apply a similar median trend analysis to a sample of 70 924 stars selected from the GALAH second data release \citep[DR2;][hereafter B18]{buder_a}. For the 13 elements in common with \citetalias{weinberg}, the GALAH measurements provide an independent check of the APOGEE median trends, derived from optical instead of near-IR spectra using a different pipeline for extracting element abundances and stellar parameters. More importantly, GALAH offers access to several new elements: C, thought to come from a combination of CCSN and asymptotic giant branch (AGB) sources; Sc and Ti, which lie between the traditional $\alpha$-elements and the elements of the Fe-peak; Cu and Zn, which lie on the steeply falling abundance trend just beyond the Fe-peak; and the neutron capture elements Y, Ba, La, and Eu. APOGEE measures C and Ti abundances,  but \citetalias{weinberg} omitted C from their study because its surface abundance in luminous giants is affected by internal evolution, and they omitted Ti because of systematic uncertainties in the APOGEE abundances. The GALAH targets are primarily main sequence stars and sub-giants, so they probe a much smaller volume than the luminous APOGEE giants. However, \citetalias{weinberg} find that median $\xmg$ vs. $\mgh$ trends are independent of location within the Galactic disk, and we can conjecture that this universality also applies to the new elements probed by GALAH. 

The 2-process decomposition method of \citetalias{weinberg} relies on the fact that stars of a given $\mgh$ span a substantial range of $\mgfe$, or more generally of $\afe$ where the ratio is based on multiple $\alpha$-elements expected to come predominantly from CCSN (e.g., O, Mg, Si, Ca). The distribution of $\afe$ depends on the stellar sample, but in the Milky Way disk it is frequently found to be bimodal, with a ``high-$\alpha$'' population dominating at large midplane distance $|Z|$ and a ``low-$\alpha$'' population with approximately solar $\afe$ dominating near the midplane \citep[e.g.,][]{bensby2003, adibekyan}. \citetalias{weinberg} interpret the plateau in the high-$\alpha$ population at $\mgfe \approx +0.3$ as representing the yields of CCSN averaged over the stellar initial mass function (IMF) and values of $\mgfe$ below this plateau as indicating additional Fe contributed by SNIa (see Section~\ref{sec:2proc} below for details). Recognizing that the physical distinction between the high-$\alpha$ and low-$\alpha$ populations is really one of Fe enhancement, \citetalias{weinberg} refer to these populations as low-$\femg$ and high-$\femg$, respectively. In this paper we adopt the more euphonious and readily interpretable terminology of ``low-Ia'' (= high-$\alpha$) and ``high-Ia'' (= low-$\alpha$), with the obvious caution that this is a theoretical interpretation of the empirical distinction.

With this interpretation, the trend of $\xmg$ vs. $\mgh$ for an element X produced entirely by CCSN should be identical for the low-Ia and high-Ia populations. For elements that are also produced by SNIa, the two populations should exhibit separated $\xmg$ vs. $\mgh$ sequences, and the greater the yield of SNIa relative to CCSN the larger the separation will be. For elements that have contributions from another source, such as AGB stars or neutron star mergers, the interpretation is more complex. If the enrichment mechanism is prompt, producing the element on the same ($\sim 10-30$ Myr) timescale as CCSN, then its trends should match in the low-Ia and high-Ia populations just as they do for CCSN production. If the mechanism produces the element over a much longer period following star formation, then the low-Ia and high-Ia populations should exhibit distinct trends, but the sequence separation will depend on both the yield of the delayed mechanism relative to CCSN and on its delay time distribution (DTD) relative to the DTD of SNIa \citep[investigated by, e.g.,][]{maoz2017}. The empirical measurement of the $\xmg$ trends is independent of these theoretical issues, but the quantitative interpretation of trend separations depends on what sources besides CCSN and SNIa contribute to a given element. 

The relative frequency of low-Ia stars is much higher at large $|Z|$ or high vertical velocity \citep{bensby2003}, leading to the often-used terminology of ``chemical thick disk'' and ``chemical thin disk'' \citep[e.g.,][]{jong2010}. Whether the observations imply a distinct origin of the thin and thick populations or merely a continuous trend of kinematics and abundance patterns with age remains a matter of debate \citep[e.g.,][]{chiappini1997, schonrich2009a, schonrich2009b, bovy2012, spitoni2019}. \citet{hayden} mapped the distribution of stars in $\afe$ vs. $\feh$ space across the Galactic disk, finding that low-Ia stars are more prevalent at small galactocentric radius $R$ as well as large $|Z|$. They also find that the metallicity distribution function (MDF) of high-Ia stars changes steadily with $R$, exhibiting both the well known disk metallicity gradient and a change of shape from skew-negative in the inner Galaxy to skew-positive in the outer Galaxy. \citetalias{weinberg} update these trends with a larger APOGEE data set, focusing on $\femg$ and $\mgh$. These population trends provide critical data for disentangling the many aspects of inflow, star formation, outflow, and radial and vertical mixing that gave rise to the present state of the Galactic disk. However, the focus of the present paper is on deriving empirical constraints on nucleosynthesis mechanisms with an approach that is relatively (though not entirely) insensitive to these other factors. 

We describe our data sample in Section~\ref{sec:methods}. In Section~\ref{sec:abundances} we derive median trends, comparing these trends to the \citetalias{weinberg} APOGEE results for elements that overlap. In Section~\ref{sec:2proc} we review the 2-process model and apply it to the GALAH median trends, again comparing to the APOGEE results for elements in common. Before fitting the GALAH trends, we apply zero-point offsets$-$a single offset per element$-$so that stars with $\femg=\mgh \approx 0$ also have $\xmg\approx 0$. We regard these generally small zero-point offsets as justified and a possible improvement to the GALAH abundance values. Section~\ref{sec:discussion} compares the results of our analysis to theoretical yield models. We summarize our conclusions in Section~\ref{sec:summary}.

\section{Data} \label{sec:methods}

To study the chemical composition of the Galaxy, we take stellar abundances from GALAH DR2 \citepalias{buder_a}. The GALAH spectroscopic survey targets stars with magnitudes $12 < V < 14$ and Galactic latitude $|b|>10$ deg, aiming to have observational overlap with \textit{Gaia} \citep[][hereafter B19]{buder_b}. Their data are composed of mainly FGK stars in the thin and thick disk, along with a substantial halo population \citepalias{buder_a}. A detailed discussion of targeting can be found in \citet{martell}. Stellar spectra are taken with the HERMES spectrograph on the 3.9-meter Anglo-Australian Telescope. HERMES observes in four optical wavelength channels with a resolution of R = 28 000 \citep{de_silva, sheinis}. \citetalias{buder_a} provide an in-depth discussion of their data reduction process. In short, they use the spectral synthesis program developed by \citet{piskunov} to derive stellar parameters for a training set of stars using a line list from \citet{heiter}. They then implement \textit{The Cannon} \citep{ness} on the entire data set. \citetalias{buder_a} report up to 23 elemental abundances for 342 682 stars.

We impose a series of quality cuts to the full GALAH population to form the sample used in this paper. GALAH returns a flag for the overall fit by \textit{The Cannon} as well as flags for each specific abundance \citepalias{buder_a}. We remove all stars with \textit{The Cannon} flag set, and we exclude stars with any elemental flags set in the analysis for that element (Section~\ref{sec:abundances}). These exclusions remove all obvious binary stars, stars with reduction issues, and elemental abundances where \textit{The Cannon} extrapolates or has a poor $\chi^2$ value. We make an overall quality cut, requiring a signal-to-noise ratio (SNR) of $> 40$ per pixel in the HERMES blue channel. 

Intrinsic abundance trends with stellar temperature or temperature dependent errors from the reduction pipeline could further affect the observed abundance medians. To investigate this issue, we compare the abundance trends of hot and cool stars to the full population. We first divide the stars roughly into thirds with the temperature groups $T_{\text{eff}} < 5 000$, $5 000 \leq T_{\text{eff}} \leq 6 000$, and $T_{\text{eff}} > 6 000$. For each element in the three groups, we calculate the median for the low-Ia and high-Ia populations and compare the trends to the entire data set. We adopt the same division as \citetalias{weinberg}, defining the low-Ia population by
\begin{equation}
    \begin{cases}
    \mgfe > 0.12 - 0.13\feh,    & \feh<0 \cr
    \mgfe > 0.12,               & \feh>0. \cr
    \end{cases}
    \label{eq:boundary}
\end{equation}

We then calculate the median absolute deviation between the median trend of a sub-sample and that of the full population to be  
\begin{equation} \label{eq:tempcut}
    \text{median}(|\text{sub-sample median} - \text{full sample median}|).
\end{equation}
We find that the middle temperature range ($5 000 \leq T_{\text{eff}} \leq 6 000$) best resembles the full population, while the hotter and cooler stars show noticeable deviations for some elements. To better isolate stars on the tail ends of the temperature distribution, we shift the divisions to $T_{\text{eff}} < 4 500$, $4 500 \leq T_{\text{eff}} \leq 6 200$, and $T_{\text{eff}} > 6 200$. These boundaries make physical sense because the lower bound removes cool dwarfs whose spectra are afflicted by molecular line blending \citepalias{buder_b} and the upper bound removes stars beyond the Kraft break \citep{kraft} whose broad lines may be poorly modeled in the GALAH analysis. With these boundaries, 80\% of stars in the high-Ia population and 87\% of stars in the low-Ia population fall within the middle range. 

\begin{figure*}[!htb]
 \includegraphics[width=\textwidth]{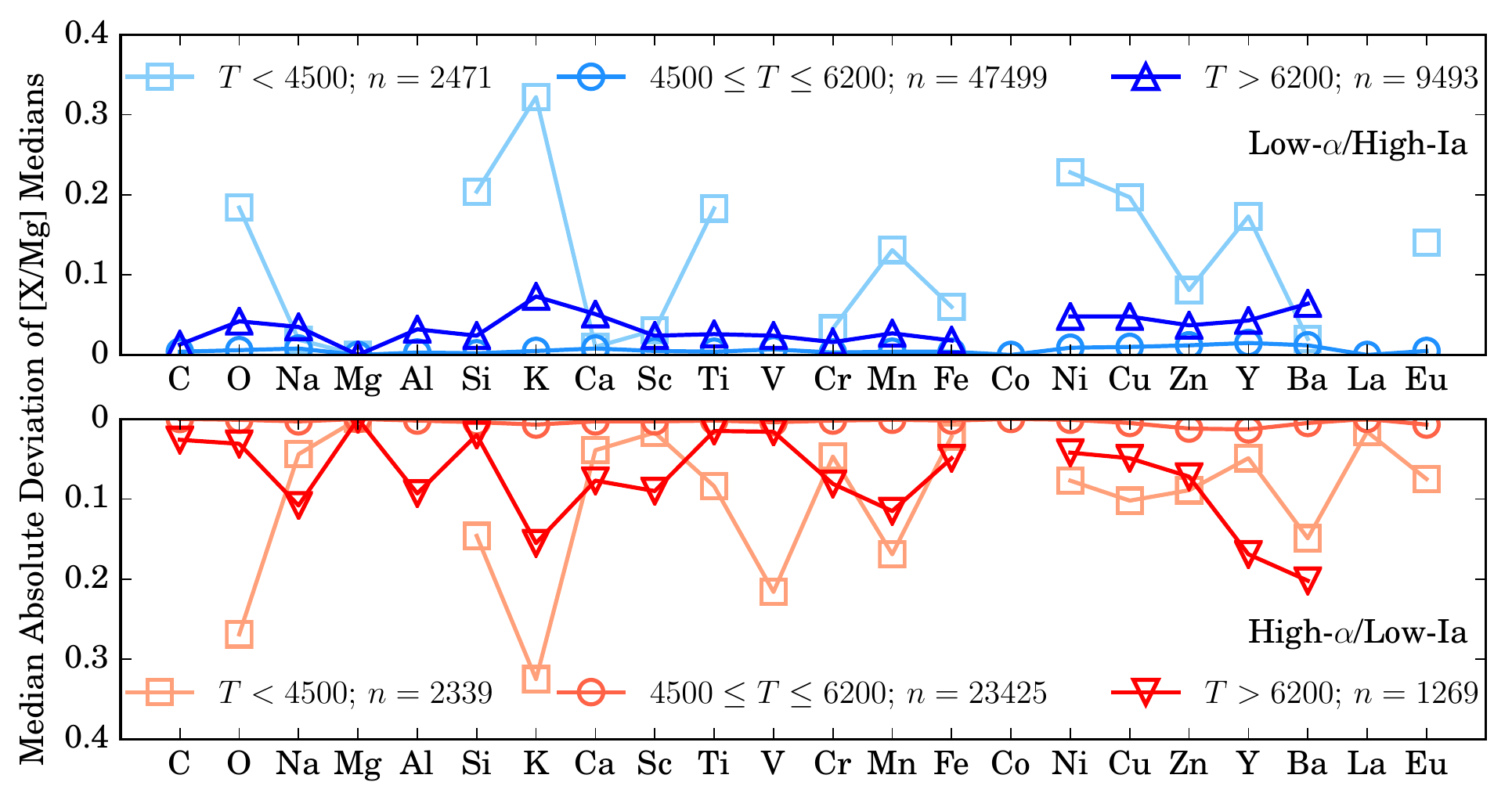}
 \caption{Deviations from the population median [X/Mg] of the low-$\alpha$/high-Ia (blue) and high-$\alpha$/low-Ia (red) sequences when stars are subdivided into three temperature groups. The median absolute deviation of median abundance ratios is defined in Equation~\ref{eq:tempcut}. The legend quotes the number of stars with unflagged Fe abundances for each temperature group. We do not plot points for elements where there are no bins with $>40$ stars. The $4500 \leq T \leq 6200$ bin tracks the total population well, with very small differences for all elements. Given the large difference between full sample and the high/low temperature groups for some elements, we remove all stars with $T_{\text{eff}}>6200$ and $T_{\text{eff}}<4500$.}
 \label{fig:temp_comp}
\end{figure*}

Figure~\ref{fig:temp_comp} plots the median absolute deviation between the $\xmg$ medians of each temperature group and the full sample medians, for the low-Ia and high-Ia populations. Elemental temperature groups for which there are no bins with $>40$ stars are not plotted. While it is not surprising that the middle temperature range is closest to the population median, the cooler stars still show deviations for many elements. Though the hotter stars better trace the full population, they too show some differences. As we have a very large stellar sample, we decided to cut both the hot and cool stars to minimize the potential impact of temperature-dependent abundance systematics on our derived element ratio trends.  

\begin{figure}[!htb]
 \includegraphics[width=\columnwidth]{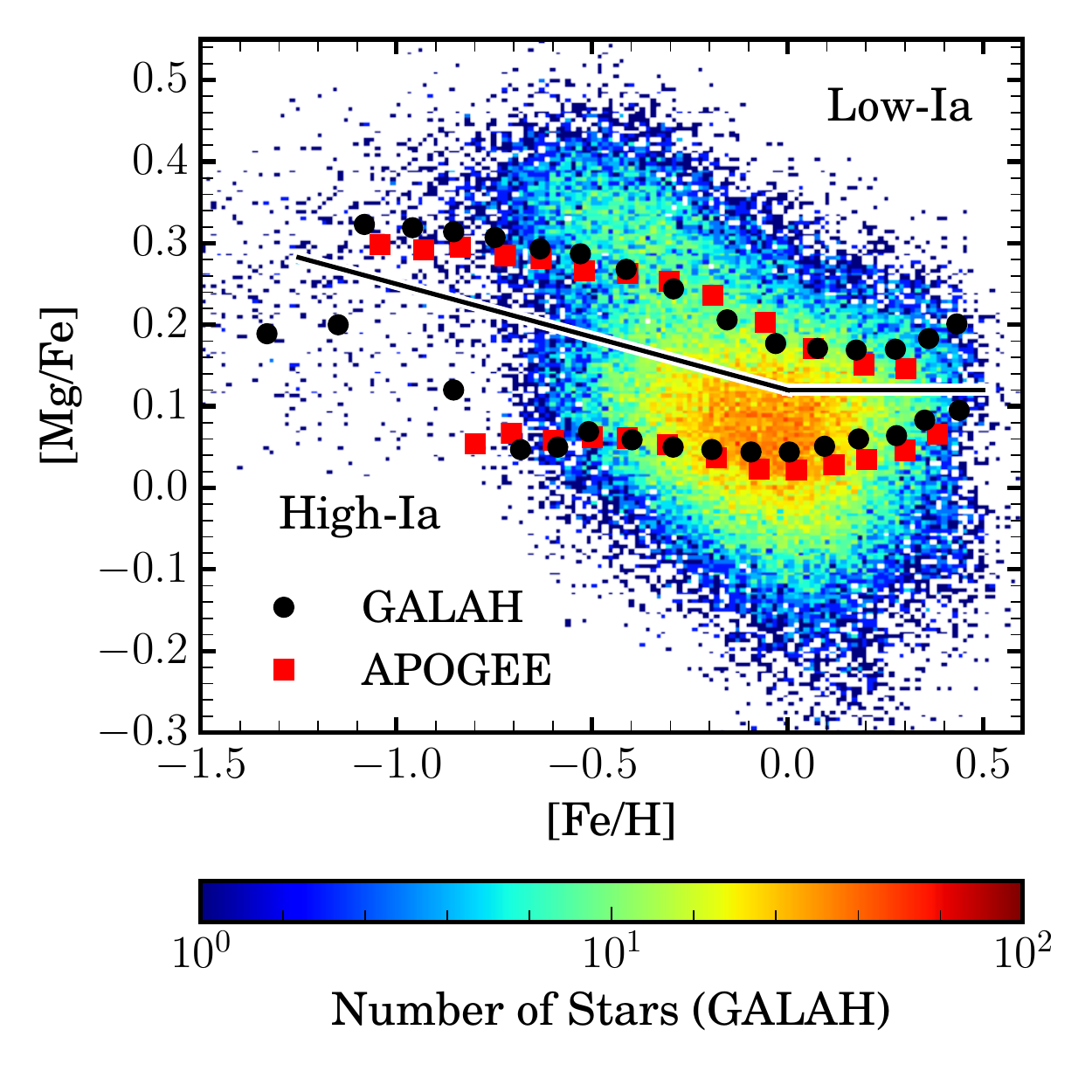}
 \caption{Distribution of 70 924 stars with SNR $\geq$ 40 and $4500\text{K} \leq T_{\text{eff}} \leq 6200\text{K}$ in [Mg/Fe] vs. [Fe/H] space. The dividing line between the high-Ia and low-Ia populations is taken from \citetalias{weinberg}. Black and red markers represent the GALAH and APOGEE median trends, respectively, for high-Ia and low-Ia populations.}
 \label{fig:star_density}
\end{figure}

Our cuts leave 70 924 stars. Figure~\ref{fig:star_density} shows their location in [Mg/Fe] vs. [Fe/H] space alongside the median values from APOGEE \citepalias{weinberg}. The bimodality between GALAH's high-Ia and low-Ia population is less defined than that of APOGEE, perhaps because of differences in relative fractions of thin and thick disk stars. Nonetheless, Equation~\ref{eq:boundary} appears to provide a reasonable separation of these populations, and once it is made then the median sequences of $\femg$ vs. $\mgh$ are very similar between APOGEE and GALAH (seen in Figures~\ref{fig:star_density} and~\ref{fig:fepeak_med}). 

As the division between high-Ia and low-Ia populations was derived for a separate survey, we test whether misclassified stars near the boundary could skew our results. For each element in the full sample, we calculate the median $\xmg$ value in $\mgh$ bins of 0.1 dex for both the high-Ia and low-Ia sequences.  We then re-calculate each, excluding stars within $\pm 0.05$ dex of the division. To compare the two samples, we find the average difference between the binned medians, defined in Equation~\ref{eq:tempcut}. We find that all elemental average differences between these two samples are less than 0.07 dex, with most around 0.03 dex. This shows that there are no major differences between median high-Ia and low-Ia sequences of the full sample and the sample excluding stars near the boundary. Therefore, possibly misclassified stars near the boundary do not skew the median abundance trends. 

\section{Stellar Abundances} \label{sec:abundances}

In the following subsections we examine the median abundance trends of stars in the GALAH catalog after implementing the quality cuts to the data described in Section~\ref{sec:methods}. Figures~\ref{fig:alpha_med} to~\ref{fig:neutron_med} plot the $\xmg$ vs. $\mgh$ trends for the high-Ia and low-Ia sequences. We plot contours at the 10th, 25th, 50th, 75th, and 90th percentiles to show the spread in the data, which comes from a combination of observational error and intrinsic scatter. Contours are not shown for endpoints whose adjacent bins have $<40$ stars. In this section, we use stellar abundances as reported by GALAH. We will include zero-point offsets in the Section~\ref{sec:2proc} analysis. Figure~\ref{fig:all_meds} in the Appendix replots all of the median trends in Figures~\ref{fig:alpha_med} to~\ref{fig:neutron_med} adjusted for these offsets. We also include median trends from APOGEE for those elements studied in \citetalias{weinberg}.

\subsection{Comparison to APOGEE} \label{subsec:overlap}

We first compare abundance trends for elements measured by both GALAH and APOGEE. Our statements about theoretical yield predictions are based mainly on the models of \citet[][hereafter AWSJ17]{andrews}, who use CCSN yields from \citet{chieffi} and \citet{limongi}, SNIa yields from \citet{iwamoto}, and AGB yields from \citet{karakas}. Results from the yield models incorporated in \textit{Chempy} \citep[][see further discussion in Section~\ref{sec:discussion}]{rybizki} are qualitatively similar. While APOGEE observes in the near-IR and GALAH in the optical, both surveys observe neutral lines for all elements except for the following: O, where APOGEE observes molecular features; Fe, where both surveys use neutral and singly ionized lines; and Cr, where GALAH uses neutral and singly ionized lines (\citetalias{buder_a}; S. Buder and K. Lind personal communication) and APOGEE uses only neutral lines.

\begin{figure}[h]
 \includegraphics[width=\columnwidth]{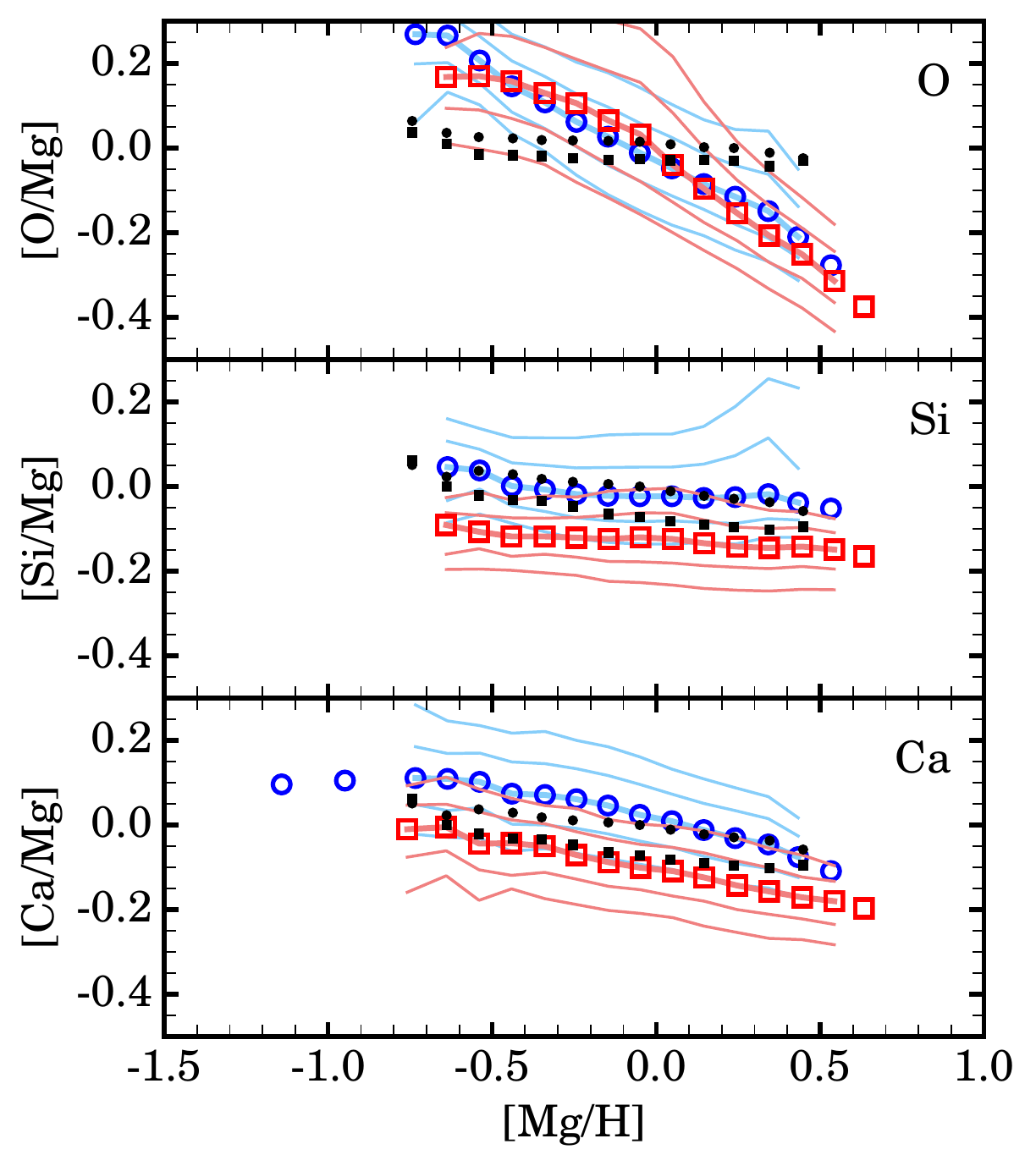}
 \caption{GALAH $\alpha$-element median abundances of the high-Ia (blue circles) and low-Ia (red squares) populations with contours at the 10th, 25th, 50th, 75th, and 90th percentiles. Data are binned by 0.1 dex in [Mg/H]. Median values are shown for bins with $>$40 stars. APOGEE median abundances from \citetalias{weinberg} are included where applicable (black markers). The GALAH and APOGEE median trends for Si and Ca are similar, but O has a much steeper metallicity dependence in GALAH.}
 \label{fig:alpha_med}
\end{figure}

\textit{$\alpha$-elements}: The GALAH and APOGEE median abundance trends for O, Si, and Ca are plotted in Figure~\ref{fig:alpha_med}. For O, trends for the low-Ia and high-Ia populations are nearly superposed in both surveys. However, the GALAH trends are strongly sloped while the APOGEE trends are nearly flat. GALAH O abundances are similar to those of \citet{bensby}, who also see a negatively sloped trend in their optically derived O abundances of F and G dwarf stars in the solar neighborhood. Close agreement of the low-Ia and high-Ia sequences is theoretically expected, as O and Mg yields are both dominated by CCSN with negligible SNIa contribution \citepalias{andrews}.

The disagreements on the metallicity dependence likely arise from the difficulty of deriving O abundances from both optical and near-IR spectra. \citetalias{buder_b} derive abundances from the O triplet, OI 7772 \AA, OI 7774 \AA, and OI 7775 \AA, lines that are strongly affected by 3D non-LTE effects \citep{kiselman, asplund, amarsi15, amarsi16}. They apply 1D non-LTE corrections in an attempt to account for these effects. The APOGEE abundances come principally from infrared molecular features (e.g. OH, CO), which are stronger and much less sensitive to non-LTE effects, though there could be other systematics in modeling these molecular features \citep[e.g.][]{collet, hayek}. The flat APOGEE trend agrees better with theoretical expectations, as O and Mg yields have only weak metallicity dependence at fixed stellar mass \citepalias[][Figure 9]{andrews}. A metallicity-dependent, IMF-averaged yield could arise if the IMF or the mass-dependence of black hole formation changes with metallicity. However, such changes would have qualitatively similar effects on O and Mg production, so it is not clear that they could cause a metallicity trend in [O/Mg].

Si and Ca median trends both show greater sequence separation than O, suggestive of increased SNIa contribution. Both also show a shallow metallicity dependence, with Ca having a steeper slope than the APOGEE medians. Ca is predicted to have a larger SNIa contribution than Si \citepalias{andrews}, but this expectation is not immediately evident in the observed high-Ia and low-Ia sequences. GALAH measures Ca with lower precision than other $\alpha$-elements and finds larger scatter in [Ca/Fe] than \citet{bensby}, though the median trends agree \citepalias{buder_b}.

\begin{figure}[h]
 \includegraphics[width=\columnwidth]{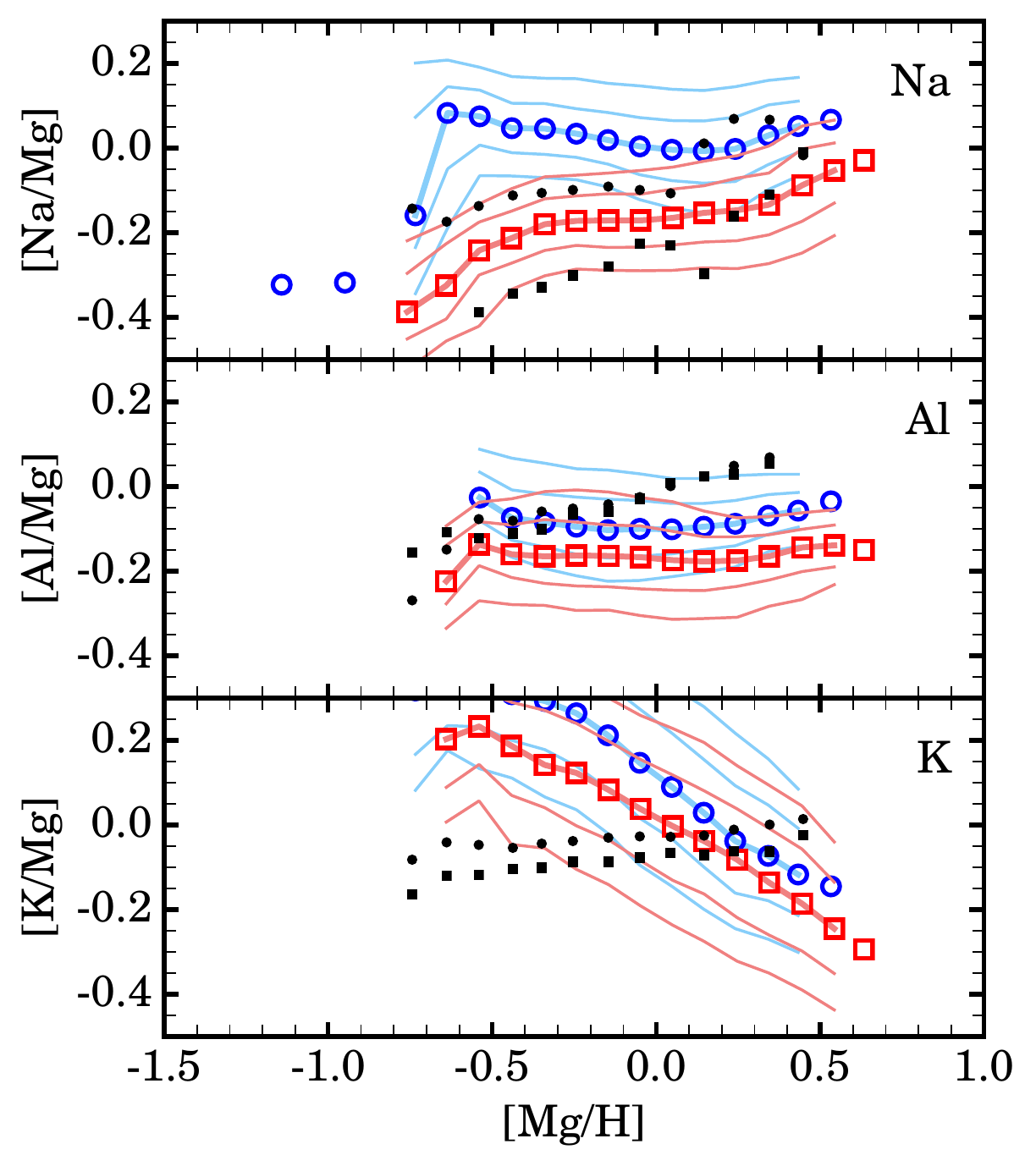}
 \caption{Same as Figure~\ref{fig:alpha_med} but for light odd-\textit{Z} elements. 
 GALAH and APOGEE [Na/Mg] trends have a strong separation between high-Ia and low-Ia sequences suggesting strong non-CCSN  contribution to Na. The metallicity trends for Al and K are different between the two surveys.}
 \label{fig:lightz_med}
\end{figure}

\textit{Light odd-Z elements}: The GALAH and APOGEE median trends for Na, Al, and K are plotted in Figure~\ref{fig:lightz_med}.  SN yield models predict these elements to be mainly produced by CCSN and have significant trends with metallicity \citepalias{andrews}. For Na, both GALAH and APOGEE show a strong separation in the trends for low-Ia and high-Ia populations. This separation suggests a large contribution from SNIa or some other non-CCSN source, in disagreement with theoretical models. The GALAH and APOGEE trends do not agree particularly well, though the difference is largely a zero-point offset, as well as an artifact at $\mgh \approx 0.15$ in the APOGEE trend discussed by \citepalias{weinberg}. The GALAH data show a strong metallicity dependence below $\mgh \approx -0.5$ and, like APOGEE, a rising [Na/Mg] with increasing $\mgh$ on the low-Ia sequence. \citetalias{buder_b} note that Na has similar behavior to the Fe peak elements Ni and Cu. 

For GALAH, the two [Al/Mg] trends show much less separation than for [Na/Mg], but more than the near-exact overlap found by APOGEE. The metallicity dependence is nearly flat, vs. weakly rising for APOGEE. Thus, while the APOGEE data suggest that Al is a pure-CCSN element with a weakly rising metallicity-dependent yield, the GALAH data suggest a moderate non-CCSN contribution and almost no metallicity dependence. The Al abundances in GALAH agree well with \citet{bensby} \citepalias{buder_b}, tracing Mg with some zero-point offset. 

GALAH and APOGEE show a comparably small separation between the high-Ia and low-Ia sequences for [K/Mg], suggesting mainly CCSN origin. However, the metallicity dependences of these trends differ drastically between the two surveys: a strong negative slope for GALAH vs. a mild positive slope for APOGEE. K suffers from strong non-LTE effects in the optical and weak lines in the near-IR, making its abundance difficult to determine. GALAH measures K from the K I 7699 $\text{\AA}$ line, which is susceptible to interstellar absorption \citepalias{buder_b}. Many theoretical yield models underpredict observed K abundances by a large factor (e.g., \citetalias{andrews}; \citealp{rybizki}; but see \citealp{tuguldur}). Figure~\ref{fig:lightz_med} shows that improvements in K abundance measurements are needed before we can draw robust conclusions about its nucleosynthetic origin.

\begin{figure*}[!htb]
\begin{center}
 \includegraphics[width=.85\textwidth]{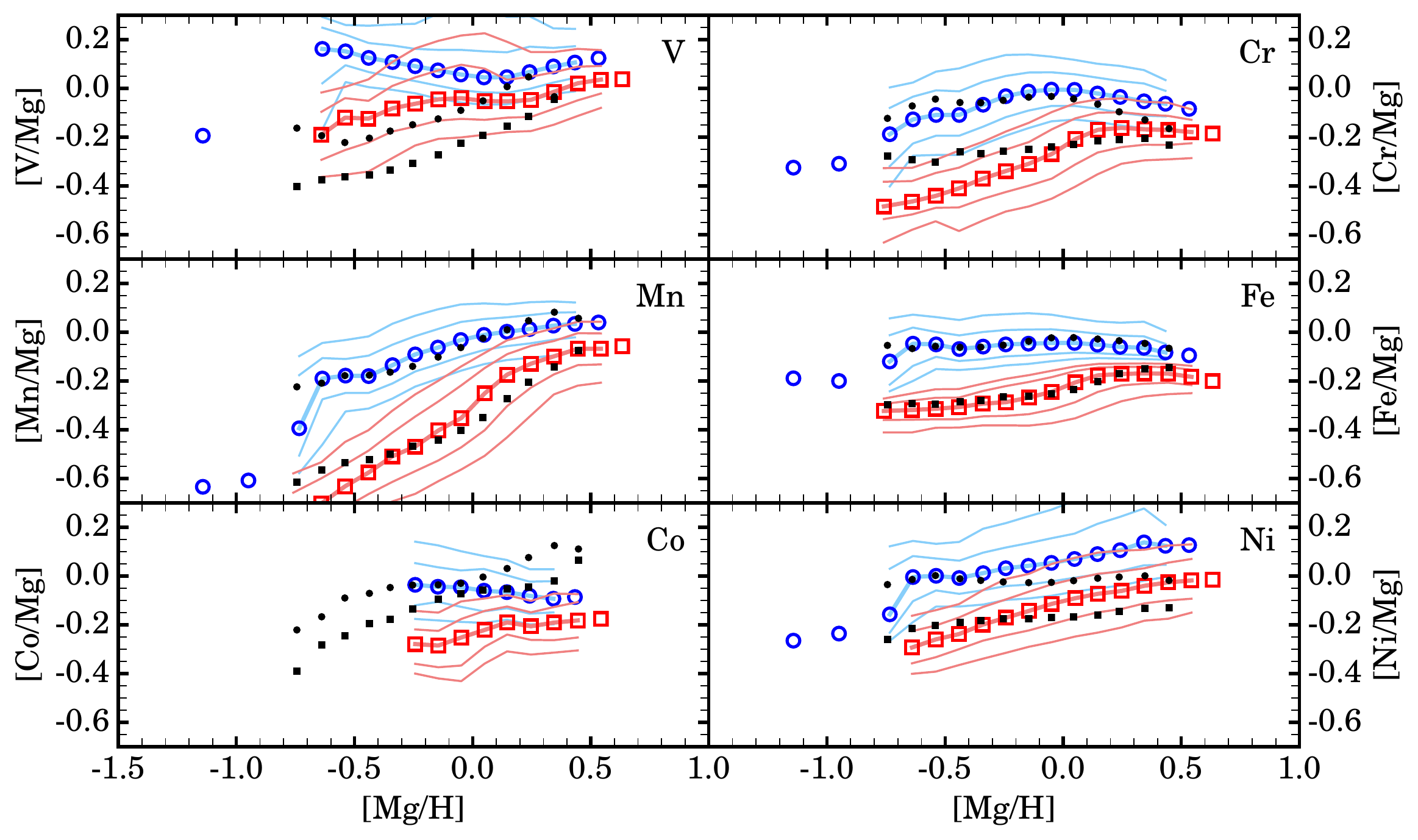}
 \caption{Same as Figure~\ref{fig:alpha_med}, but for Fe peak elements with the y-axis range expanded. Odd-\textit{Z} elements are in the left hand column and even-\textit{Z} elements are on the right.
 All show some separation between the high-Ia and low-Ia sequences, indicative of combined CCSN and SNIa contribution. GALAH's Mn, Cr, Fe, and Ni trends agree reasonably well with APOGEE's while V and Co differ in metallicity dependence.}
 \label{fig:fepeak_med}
 \end{center}
\end{figure*}

\textit{Fe-peak elements}: Median trends for odd-\textit{Z} elements (V, Mn, and Co) are plotted in the left hand column of Figure~\ref{fig:fepeak_med} and even-\textit{Z} elements (Cr, Fe, and Ni) are plotted on the right. 
We expect all Fe-peak elements included here to have both CCSN and SNIa contribution. SN yields for odd-\textit{Z} elements also predict positive metallicity dependence \citepalias{andrews}. We see obvious separations between the high-Ia and low-Ia sequences in [V/Mg], [Mn/Mg], and [Co/Mg]. 
The [V/Mg] values are higher in GALAH than APOGEE, with a flatter trend. The V lines employed in both surveys are weak and susceptible to blending in metal-rich and metal-poor stars, so these trends should be viewed with caution (\citealp{jonsson}; \citetalias{buder_b}). 

Mn shows the most separation between the high-Ia and low-Ia sequences, supporting yield predictions that it has the largest SNIa contribution of the odd-\textit{Z} elements shown here \citepalias{andrews}. Mn abundances agree well between GALAH and APOGEE. Both surveys assume LTE when deriving Mn abundances. The observed metallicity dependence may flatten after applying non-LTE corrections \citep{bensby_Eu}.
Co was only detected for $\sim 4 \%$ of stars observed by GALAH \citepalias{buder_b}. Its metallicity dependence in GALAH is much flatter than in APOGEE. 

Even-\textit{Z} elements display weaker metallicity dependences than the odd-\textit{Z} elements. We note that the high-Ia and low-Ia populations for Fe are perfectly separated at all percentile levels by definition (Equation~\ref{eq:boundary}). As in APOGEE, the GALAH low-Ia median plateaus around $\femg$ of -0.3 dex. [Cr/Mg] trends are roughly similar to $\femg$ indicative of similar production processes. The GALAH [Ni/Mg] trends differ from APOGEE, showing a steeper metallicity dependence and potential zero-point offset.

\subsection{New elements}\label{subsec:new}

GALAH DR2 includes elemental abundances that are not included in APOGEE DR14, with Sc, Ti, Cu, Zn, Y, Ba, La, and Eu for a large sample of stars, and C for dwarf stars with $\mgh \gtrsim -0.3$. APOGEE measures C for all stars, but in the luminous giants used by \citetalias{weinberg} the C abundances are affected by internal production and dredge-up, so they cannot be interpreted as birth abundances.

\begin{figure}[h]
 \includegraphics[width=\columnwidth]{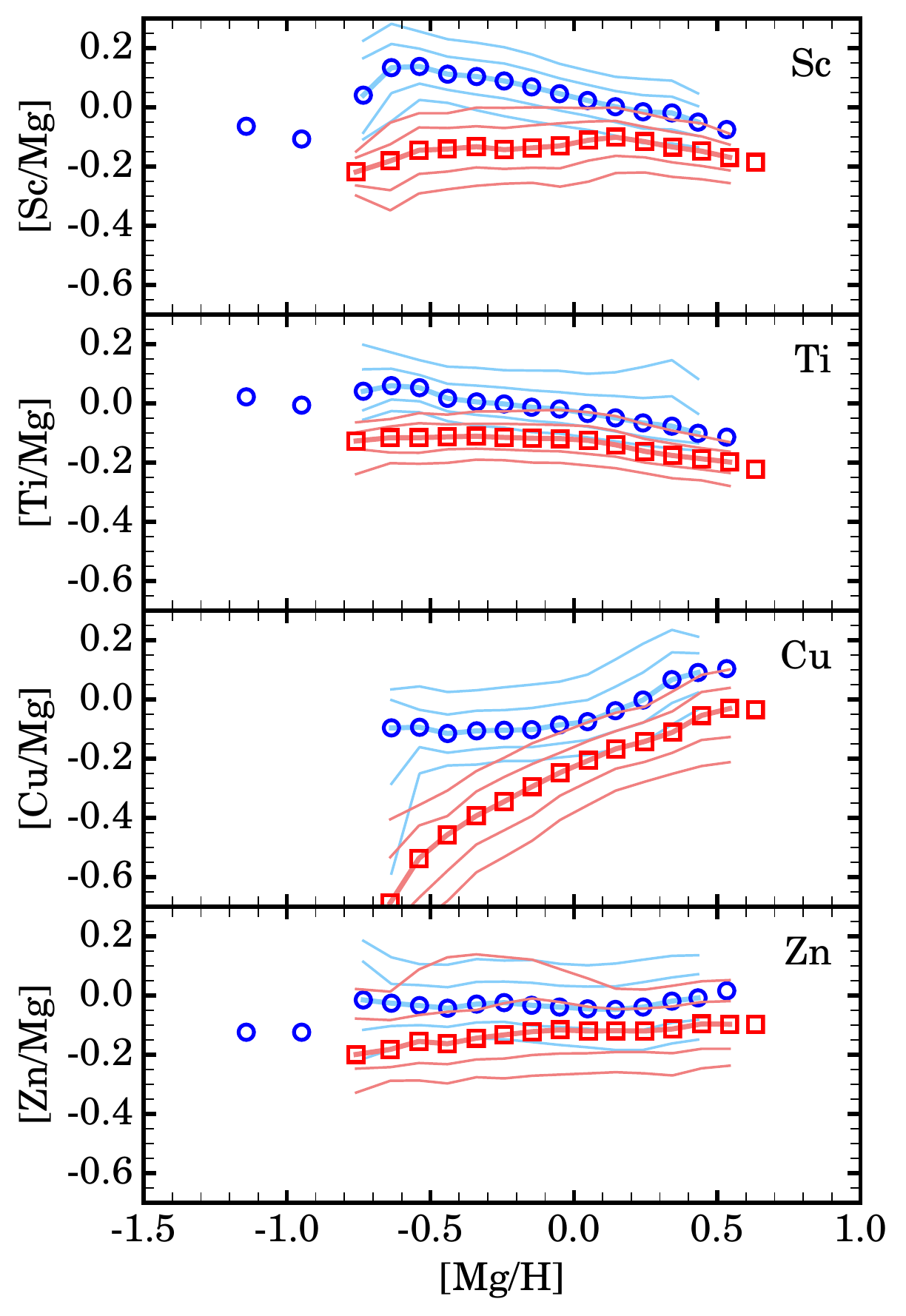}
 \caption{Same as Figure~\ref{fig:alpha_med}, but for new Fe-peak and Fe-cliff elements in GALAH with the y-axis range expanded. The separation between the low-Ia and high-Ia sequences suggest that all four elements have CCSN and SNIa contributions. Cu shows steeply increasing yields while the other trends are relatively metallicity independent.
 }
 \label{fig:newsn_med}
\end{figure}

\textit{Fe-peak and Fe-cliff}:  The median trends of elements Sc, Ti, Cu, and Zn are plotted in Figure~\ref{fig:newsn_med}. Here and in later sections, we label Cu and Zn as ``Fe-cliff'' because they are on the steeply falling edge of the Fe-peak and may have distinct nucleosynthesis from elements shown in Figure~\ref{fig:fepeak_med}. Ti, an even-\textit{Z} element categorized as both an $\alpha$-element and Fe-peak, has both CCSN and SNIa contribution \citepalias{andrews}. GALAH measures Ti very well, as the species has many clean lines in the optical. The GALAH [Ti/Mg] abundance trends resemble $\alpha$-elements Si and Ca. Both the high-Ia and low-Ia sequences are flat and have minor separation, suggesting mainly CCSN origin with little metallicity dependence. GALAH Ti trends agree well with measurements from \citet{bensby} \citepalias{buder_b}.

[Sc/Mg] appears similar to [Fe/Mg] and [Cr/Mg], with diverging high-Ia and low-Ia sequence at low [Mg/H] but flatter, concurrent trends at higher metallicity. We see some separation between the sequences, suggesting SNIa contribution, and a negative metallicity dependence. \citetalias{buder_b} note similar trends in [Sc/Fe] space to those of \citet{bensby_Sc}.

The median trends for [Cu/Mg] are different from many of the previous Fe-peak elements, but most resemble those of Mn. We see a strong, positive metallicity dependence and a separation between the high-Ia and low-Ia sequences. This separation appears to increase at low metallicity. Median trends do not extend to low [Mg/H], as the lines used to derive Cu abundances in GALAH spectra are not detected in metal poor stars \citepalias{buder_b}. GALAH Cu abundances agree well with those of \citet{delgado}, a study of 1111 FGK stars in the HARPS GTO sample \citepalias{buder_b}.  

Like Ti, the [Zn/Mg] median trends also resemble an $\alpha$-element with a small SNIa contribution. We see a mild separation between the high-Ia and low-Ia sequences and no metallicity dependence. Zn abundance derivations from cool stars in GALAH suffer from effects of line blending, causing lower precision in the measurements for this element \citepalias{buder_b}. Data from \citet{bensby} lie in a similar but offset region of [Zn/Fe] space \citepalias{buder_b}.

The yield models in \citetalias{andrews} predict negligible SNIa contribution to Sc, Cu, and Zn, but the separations observed in the GALAH high-Ia and low-Ia sequences suggest a moderate contribution from SNIa or some other delayed component. 

\textit{Carbon}: GALAH measures C abundances from atomic lines with high excitation energies. As these lines are only visible in hot, metal rich stars \citepalias{buder_b}, the [C/Mg] abundances do not extend below [Mg/H] of -0.4 dex and are limited to dwarf stars. We see some separation between the high-Ia and low-Ia sequences indicative of a non-CCSN contribution. In the case of C, this delayed contribution may be associated with AGB stars rather than SNIa \citepalias{andrews}. The observed GALAH sequences show a strong, decreasing metallicity trend over the range $-0.4 \leq \mgh \leq 0.5$.

\begin{figure}[h]
 \includegraphics[width=\columnwidth]{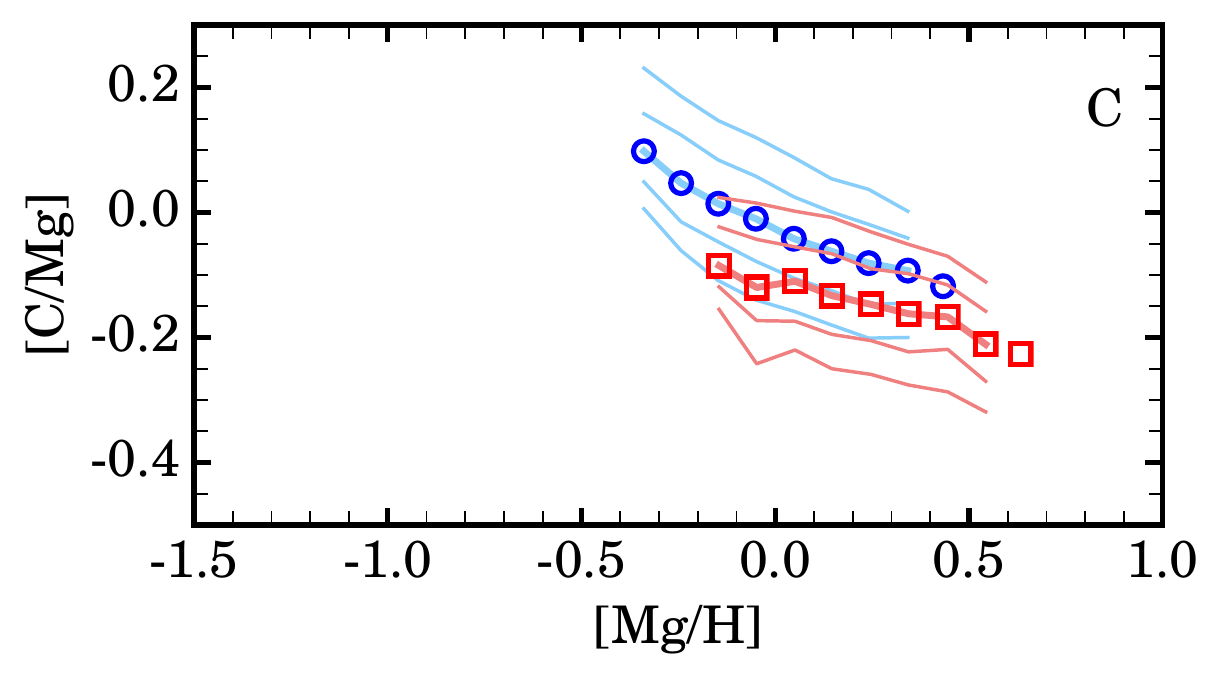}
 \caption{Same as Figure~\ref{fig:alpha_med}, but for Carbon.
 The high-Ia and low-Ia sequences are slightly separated and negatively sloped, suggesting metallicity dependent CCSN and SNIa contribution.}
 \label{fig:carbon_med}
\end{figure}

\begin{figure*}[!ht]
\begin{center}
 \includegraphics[width=.85\textwidth]{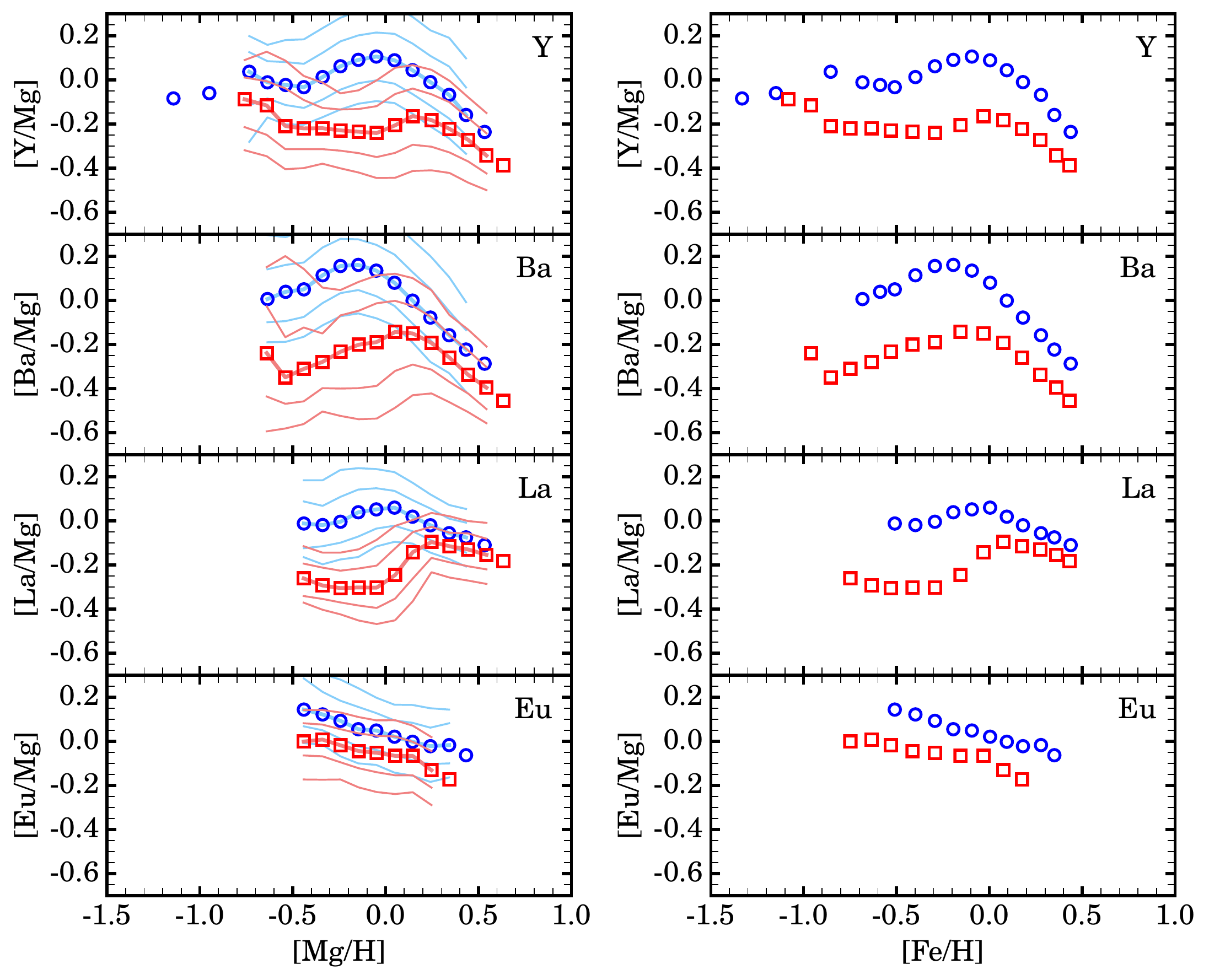}
 \caption{Left: Same as Figure~\ref{fig:alpha_med} but for neutron capture elements with the y-axis range expanded.
All neutron capture elements show some separation between the high-Ia and low-Ia sequences, suggesting combined contribution from a prompt and a delayed process. All also have metallicity dependent trends, with Y, Ba, and La's taking a non-linear form. Right: Median high-Ia and low-Ia sequences for neutron capture elements plotted against $\feh$. The peaks in median sequences are offset in $\mgh$ but line up in $\feh$.}
 \label{fig:neutron_med}
 \end{center}
\end{figure*}

\textit{Neutron capture elements}: GALAH further includes abundances for high-Z elements Y, Ba, La, and Eu. We plot the median abundance trends for these elements in the left-hand panel of Figure~\ref{fig:neutron_med}. A combination of the slow and rapid neutron capture processes ($s$- and $r$-process) produces Y, Ba, and La, with Ba and Y having a mostly $s$-process origin in the solar system \citep{arlandini,bisterzo}. Recent works have quantified the fractional contribution of each process \citep{bensby_Eu}, but suggest an additional primary production process that acts at low-metallicities \citep{travaglio}.  

In Figure~\ref{fig:neutron_med}, Y, Ba, and La show substantial separation between the high-Ia and low-Ia sequences, suggesting contribution from prompt and delayed production mechanisms. As with C, the delayed production mechanism for these elements is likely to be associated with AGB stars rather than SNIa. These three elements also have non-linear metallicity dependences, exhibiting a turn over at [Mg/H] around solar. The peak in the low-Ia stars falls at a higher value of [Mg/H] than the peak in the high-Ia stars. To understand this offset, we include a similar plot, but with Fe as the reference element in the right-hand panel of Figure~\ref{fig:neutron_med}. Here the high-Ia and low-Ia peaks in Y, Ba, and La abundance occur at about the same values of $\feh$. The similar sequence behavior with respect to Fe supports the view that Fe and/or Fe-peak elements are providing the seed nuclei for formation of these neutron capture elements, as discussed in \citet{kappeler}. Correspondingly, the amount of Y, Ba, and La depends on the Fe abundances and not the Mg abundance. Theoretically, the Ba yield increases with [Fe/H] at low metallicity because the number of seeds for neutron capture increases, then turns over at high [Fe/H] because the number of neutrons per seed is too small to produce elements as heavy as Ba \citep{gallino}.

\begin{figure*}[!ht]
\begin{center}
 \includegraphics[width=.77\textwidth]{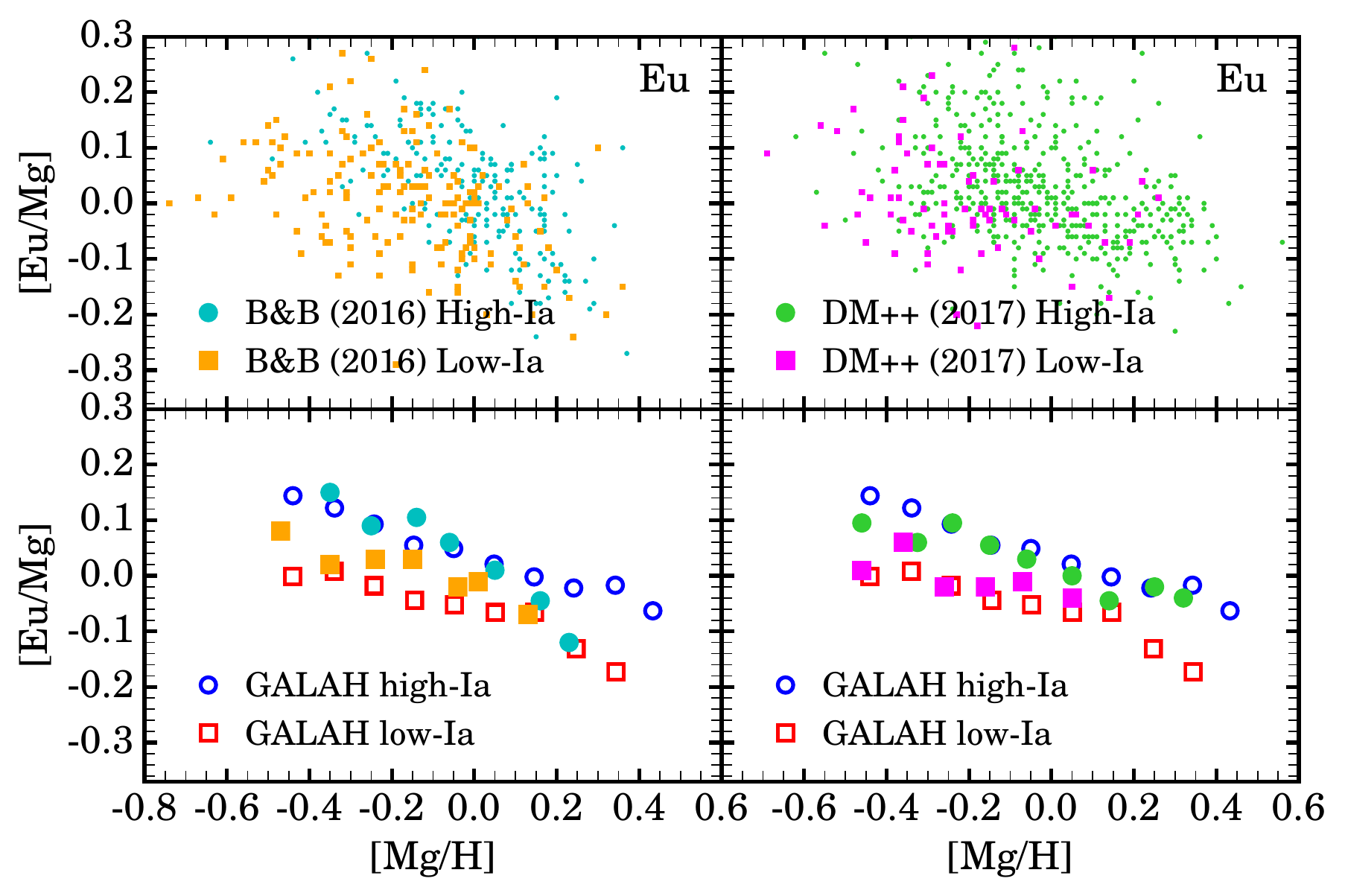}
 \caption{Top Left: Distribution of Eu abundances in 378 dwarf stars from \citet{bensby_Eu}. Stars classified as high-Ia and low-Ia (Equation~\ref{eq:boundary}) are colored teal and orange, respectively.
 Top Right: Distribution of Eu abundances in 570 FGK stars from \citet{delgado} with green high-Ia and magenta low-Ia stars. Bottom Left and Right: Median abundances of the high-Ia (filled teal/green circles) and low-Ia (filled orange/magenta squares) populations. GALAH median trends (bottom left panel of Figure~\ref{fig:neutron_med}) are added for comparison.
 }
 \label{fig:bensby}
 \end{center}
\end{figure*}

Eu differs from the other neutron capture elements included in GALAH as the $r$-process dominates its production \citep{arlandini,bensby_Eu}. We see some separation between the high-Ia and low-Ia sequences in Figure~\ref{fig:neutron_med}, suggesting that some Eu is produced by a time-delayed source. \citet{bensby_Eu} include Eu and Mg in their study of dwarf abundances and \citet{delgado} in their study of FGK stars in the HARPS GTO sample. We derived the median trends of the high-Ia and low-Ia stars in both data sets, each of which contain less than 1000 stars.  Both samples and their comparison to GALAH are shown in Figure~\ref{fig:bensby}. Although the sequence separations are smaller than in GALAH, these samples are too small to clearly confirm or clearly contradict the GALAH result. All show a linearly declining metallicity dependence.

\section{2-process model} \label{sec:2proc}
To quantify the SNIa contribution and the metallicity dependence of the CCSN and SNIa yields for elements observed by GALAH, we employ the 2-process model developed by \citetalias{weinberg}. The model describes all of a star's elemental abundances as the sum of a prompt CCSN process, $\pcc$, and a delayed SNIa process, $\pIa$, with amplitudes $\Acc$ and $\AIa$, respectively. It defines the ratio of these processes for some element, X, as 
\begin{equation}  \label{eq:ria}
    \RIa \equiv \frac{p_{\rm Ia, \odot}^{\rm X}}{p_{\rm cc, \odot}^{\rm X}},
\end{equation}
where $p_{\rm Ia, \odot}^{\rm X}$ and $p_{\rm cc, \odot}^{\rm X}$ are the process contributions at solar metallicity (defined by $\mgh=0$), whose sum returns the solar abundance of element X. Both processes are allowed some metallicity dependence, modeled as power laws with slopes $\acc$ and $\aIa$ such that
\begin{equation} \label{eq:pcc}
\pcc(Z) = p_{\rm cc, \odot}^{\rm X} \cdot10^{\acc\mgh}
\end{equation}and
\begin{equation} \label{eq:pia}
\pIa (Z) = p_{\rm Ia, \odot}^{\rm X} \cdot 10^{\aIa\mgh} \, .
\end{equation}
The parameters $\RIa$, $\acc$, and $\aIa$ are fixed for each element $X$, while the parameters $\Acc$ and $\AIa$ vary from star to star.

The 2-process model is based on three main assumptions: CCSN produce all Mg with metallicity independent yields, CCSN and SNIa both produce Fe with metallicity independent yields, and stars on the low-Ia plateau have purely CCSN enrichment. The observed level of this plateau at $\mgfe\approx +0.3 \approx \log(2)$ implies that the CCSN and SNIa contributions to Fe are equal at $\mgfe=0$. For a given star, the implied ratio of SNIa and CCSN amplitudes is
\begin{equation} \label{eq:aratio}
	\frac{\AIa}{\Acc} = 10^{0.3 - \mgfe} -1 \, .
\end{equation}
By construction, Equation~\ref{eq:aratio} returns $\AIa/\Acc \approx 1$ for a star of solar $\mgfe$ and $\AIa/\Acc = 0$ for a star on the low-Ia plateau.

Following the derivation in \citetalias{weinberg}, the expected abundance of element X relative to Mg can be expressed in terms of the global model parameters ($\RIa$, $\acc$, and $\aIa$) and a star's measured Mg and Fe abundances as
\begin{multline} \label{eq:xmg}
	\xmg = \acc \mgh + \\
	\log \bigg[ \frac{1+\RIa (\AIa/\Acc)  10^{(\aIa-\acc)\mgh} }{1 + \RIa} \bigg], 
\end{multline}
with $\AIa/\Acc$ inferred from Equation~\ref{eq:aratio}. Although Equation~\ref{eq:xmg} can be applied to individual stars, in this paper (as in \citetalias{weinberg}) we apply it to the median $\xmg$ ratios of the low-Ia and high-Ia populations, with the goal of inferring the values of $\RIa$, $\acc$, and $\aIa$ for the GALAH elements. For elements for which the dominant non-CCSN production is likely to come from AGB stars rather than SNIa (e.g., C, Y, Ba, La, and maybe Na and P), the model parameters should be regarded as only qualitative indications, since the time profile of AGB enrichment will not match that of SNIa.

To derive the best $\RIa, \acc,$ and $\aIa$ values for our population, we perform an unweighted, least-squares fit of the 2-process model. We simultaneously fit the high-Ia and low-Ia median sequences of each element, requiring the same parameters for both populations. Because the model is almost certainly too simple to describe these sequences within the tiny statistical errors of the median ratios, a weighted fit is less appropriate and formal $\chi^2$ values are not meaningful. As in \citetalias{weinberg}, we conduct a grid search for each free-parameter with a grid step size of 0.01. We run two fits, one restricting $\aIa=0$, and one with all three parameters free. We find similar fit quality and $\RIa$ values for both, so we report only the three free-parameter model here. Given the tiny statistical errors, we do not report error bars on the fit parameters, as they would be small but not meaningful. 

Qualitatively, a larger $\RIa$ value drives up the SNIa contribution and increases the separation between the high-Ia and low-Ia sequences. Positive $\acc$ and $\aIa$ values correspond to increasing metallicity dependence, and negative to decreasing metallicity dependence. $\acc$ tilts both sequences, while $\aIa$ can change the relative tilt of the the high-Ia sequence. As the model assumes that Mg yields have no metallicity dependence, the $\acc$ and $\aIa$ parameters really represent the metallicity dependences relative to that of Mg. 

In Figures~\ref{fig:alpha_med} to~\ref{fig:neutron_med} we use GALAH abundance ratios as reported in DR2. However, global zero-point offsets, like those applied to APOGEE data by \citet{holtzman}, are quite plausible given the inevitable imperfections of abundance determinations. The observed high-Ia sequence has $\femg \approx 0$ at $\mgh=0$, and one can see from Equations~\ref{eq:aratio} and~\ref{eq:xmg} that the 2-process model therefore predicts $\xmg=0$ at $\mgh=0$ along this sequence regardless of the values of $\RIa, \acc,$ and $\aIa$. The model cannot obtain a good fit to a high-Ia sequence that does not have $\xmg=0$ at $\mgh=0$, and a zero-point calibration error may distort fitted parameter values. We have therefore chosen to apply a global zero-point offset for each GALAH element, to both the low-Ia and high-Ia sequences, such that each high-Ia sequence has $\xmg=0$ at $\mgh=0$. (\citetalias{weinberg} did not do this because the \citet{holtzman} calibration offsets already enforced this condition to a good approximation.) We report these offsets in Table~\ref{tab:zeros}; they have an average value of 0.08 dex and a maximum of 0.12 dex, similar in scale to those applied by \citet{holtzman}. We regard these offsets as plausible corrections to the new GALAH abundances. Our $+0.044$ dex offset for $\femg$ forces $\femg=0$ at $\mgh=0$ on the high-Ia sequence and affects all fit parameters because of the influence of $\mgfe$ in Equation~\ref{eq:aratio}. In the Appendix, Figure~\ref{fig:all_meds} plots all GALAH median sequences including these zero-point offsets. Tables~\ref{tab:overlap_meds} and~\ref{tab:new_meds} list median sequences including these offsets.

Figures~\ref{fig:zeros_2proc} to~\ref{fig:neutron_2proc} show the best fit 2-process model and corresponding parameters alongside the median GALAH trends. We also quote model parameters in Table~\ref{tab:zeros}. Where applicable, we include the APOGEE 2-process model curves and parameters in the figures. 

\begin{deluxetable}{lrrrrrr}[h]
\tablecaption{Number of stars, zero-point offsets, and best fit 2-process model parameters for each element. $\fcc$ denotes the fractional CCSN contribution, as defined in Equation~\ref{eq:fcc}.\label{tab:zeros}}
\tablehead{
\colhead{[X/Mg]} & \colhead{$N_*$} & \colhead{Offset} & \colhead{$\RIa$} & \colhead{$\acc$} & \colhead{$\aIa$} & \colhead{$\fcc$}
}
\startdata
C & 12 381    & $0.026$ & $ 0.36 $ &  $ -0.33 $ & $  0.20 $ & $ 0.74 $ \\
O & 67 362    & $0.030$ & $ 0.00 $ &  $ -0.40 $ & $  0.00 $ & $ 1.00 $\\
Na & 68 926   & $-0.001$ & $ 0.72 $ &  $  0.15 $ & $ -0.30 $ & $ 0.58 $\\
Al & 43 654   & $0.101$ & $ 0.35 $ &  $ -0.20 $ & $  0.70 $ & $ 0.74 $\\
Si & 67 188   & $0.023$ & $ 0.52 $ &  $ -0.20 $ & $  0.28 $ & $ 0.66 $\\
K & 55 083    & $-0.118$ & $ 0.41 $ &  $ -0.40 $ & $ -0.40 $ & $ 0.71 $\\
Ca & 63 895   & $-0.016$ & $ 0.57 $ &  $ -0.25 $ & $ -0.02 $ & $ 0.64 $\\
Sc & 66 616   & $-0.034$ & $ 0.79 $ &  $ -0.03 $ & $ -0.34 $ & $ 0.56 $\\
Ti & 63 017   & $0.027$  & $ 0.46 $ &  $ -0.11 $ & $ -0.24 $ & $ 0.69 $\\
V & 49 654    & $-0.051$ & $ 0.43 $ &  $  0.09 $ & $ -0.40 $ & $ 0.70 $\\
Cr & 62 751   & $0.006$ & $ 1.72 $ &  $  0.08 $ & $  0.11 $ & $ 0.39 $\\
Mn & 62 887   & $0.021$ & $ 2.05 $ &  $  0.39 $ & $  0.23 $ & $ 0.33 $\\
Fe & 70 924   & $0.044$ & $ 0.99 $ &  $  0.00 $ & $  0.00 $ & $ 0.50 $\\
Co & 2 752    & $0.053$  & $ 1.01 $ &  $ -0.04 $ & $ -0.09 $ & $ 0.50 $\\
Ni & 62 585   & $-0.062$ & $ 0.93 $ &  $  0.10 $ & $  0.16 $ & $ 0.52 $\\
Cu & 40 861   & $0.080$ & $ 0.71 $ &  $  0.56 $ & $ -0.40 $ & $ 0.59 $\\
Zn & 67 686   & $0.042$ & $ 0.29 $ &  $  0.06 $ & $ -0.34 $ & $ 0.78 $\\
Y & 66 285    & $-0.097$ & $ 2.63 $ &  $ -0.40 $ & $  0.08 $ & $ 0.28 $\\
Ba & 48 859   & $-0.101$ & $ 3.21 $ &  $ -0.29 $ & $ -0.16 $ & $ 0.24 $\\
La & 17 245   & $-0.056$ & $ 2.31 $ &  $ -0.23 $ & $  0.09 $ & $ 0.30 $\\
Eu & 11 378   & $-0.035$ & $ 0.42 $ &  $ -0.27 $ & $ -0.26 $ & $ 0.70 $\\
\enddata
\end{deluxetable}

\subsection{Light-\textit{Z} elements}\label{subsec:2proc_light}

\begin{figure*}[!htb]
\begin{center}
 \includegraphics[width=.85\textwidth]{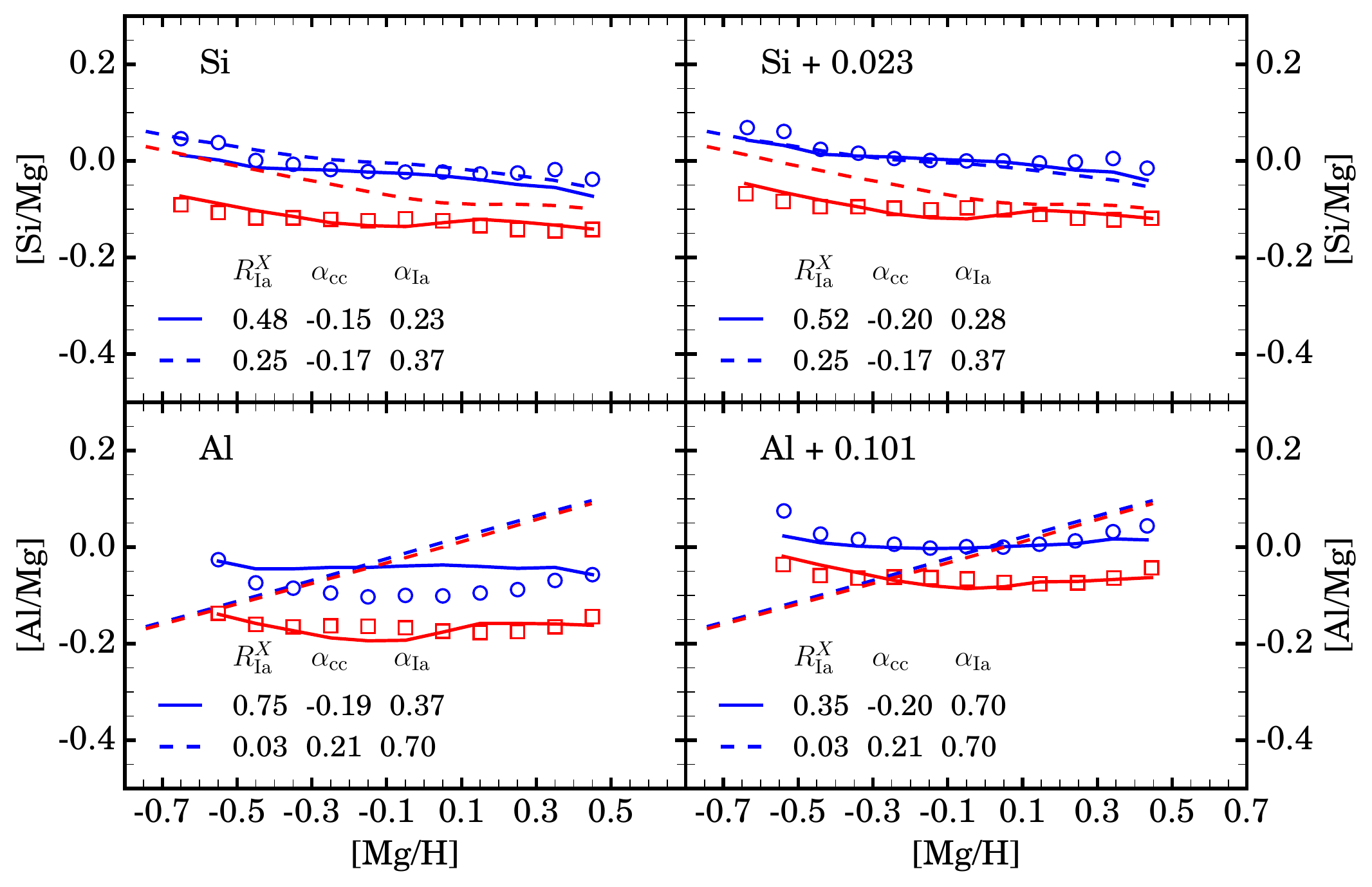}
 \caption{Median sequences and the 2-process model for for Si and Al, before (left-hand column) and after (right-hand column) zero-point offsets. Blue circles and red squares mark the median GALAH abundances, binned by 0.1 dex, for the high-Ia and low-Ia sequences, respectively. The solid lines represents the best-fit 2-process model to the GALAH data. The dashed lines (the same in both columns) are the best-fit 2-process model from APOGEE \citepalias{weinberg}. GALAH and APOGEE 2-process model parameters for the relative SNIa to CCSN contribution ($\RIa$), and CCSN ($\acc$) and SNIa ($\aIa$) metallicity dependence are included in each cell along with the zero-point offset, where applicable. We see improvement in both fits after the zero-point corrections, though more evident for Al.}
 \label{fig:zeros_2proc}
 \end{center}
\end{figure*}

Figure~\ref{fig:zeros_2proc} illustrates the 2-process fits and the impact of our adopted zero-point offsets for two representative elements, Si and Al. For Si the zero-point offset is small (+0.023 dex), and it has little impact on the visual quality of the fit or on the inferred values of $\RIa, \acc,$ and $\aIa$ (compare top left and top right panels). The significant separation of [Si/Mg] between the low-Ia and high-Ia populations implies a significant SNIa contribution to Si: the fitted value of $\RIa = 0.52$ implies that SNIa contribute about 1/3 of the Si to stars with $\mgh=\mgfe=0$. The downward slope of the trends at low $\mgh$ implies a negative $\acc\approx-0.2$, while the positive inferred $\aIa\approx0.3$ flattens the predicted trend at high $\mgh$.

For Al the +0.101 zero-point offset makes an important difference to the fit quality and to inferred parameter values. With no offset (lower left panel) the model cannot fit the measured sub-solar [Al/Mg] of the high-Ia sequence, and a large value of $\RIa=0.75$ is needed to fit the low-Ia sequence. With the offset (lower right panel) the model fits both sequences well with a smaller $\RIa=0.35$. The fitted value of $\aIa=0.70$ is large, but we caution that $\aIa$ values are poorly constrained if $\RIa$ itself is small, since the SNIa contribution then has only a small impact on the predicted $\xmg$ ratios at any metallicity. 

Dashed curves show the 2-process model fits to APOGEE data from \citetalias{weinberg}. These curves trace the APOGEE median sequences quite well for most elements (see Figures 14-16 of \citetalias{weinberg}), so differences between dashed and solid curves mainly reflect differences in the observed median sequences, as one can confirm from our Figures~\ref{fig:alpha_med} to~\ref{fig:fepeak_med} and Figure~\ref{fig:all_meds}. As noted previously in Section~\ref{sec:abundances}, the [Si/Mg] trends agree fairly well between APOGEE and GALAH, though the APOGEE [Si/Mg] ratios are higher for the low-Ia sequence at low metallicity. For Al, on the other hand, the trends from the two surveys are quite different, with APOGEE showing nearly superposed low-Ia and high-Ia sequences with a positive slope in contrast to separated sequences that are approximately flat. The APOGEE measurements and inferred model parameters agree better with theoretical expectations that Al is a nearly pure-CCSN element with a positive dependence of yield on metallicity because it is an odd-\textit{Z} nucleus. 

It may seem surprising that the GALAH and APOGEE best-fit models predict similar values of [Si/Mg] at [Mg/H]=0 despite their quite different values of $\RIa$ (0.52 vs. 0.25). However, the observed $\femg-\mgh$ sequences, though visually similar (see Figures~\ref{fig:fepeak_med} and~\ref{fig:fepeak_2proc}), are different enough between the two surveys to affect the model predictions, especially at the high metallicity end of the low-Ia sequence. In particular, the median $\femg$ values at $\mgh=0$ imply (through Equation~\ref{eq:aratio}) $\AIa/\Acc=0.32$ for GALAH but only 0.14 for APOGEE. For $\mgh=0$, Equation~\ref{eq:xmg} can be simplified and inverted to yield
\begin{equation}
    \RIa = (1-10^{\xmg})/(10^{\xmg} - \AIa/\Acc).
\end{equation}
The inferred value of $\RIa$ is higher for GALAH both because $10^{\xmg}$ is lower (0.80 vs. 0.84), raising the numerator, and because $\AIa/\Acc$ is higher, reducing the denominator. The cautionary takeaway is that even moderate differences in the observed abundance trends can have significant impact on the inferred parameters of the 2-process model. 

\begin{figure}[h]
\begin{center}
 \includegraphics[width=\columnwidth]{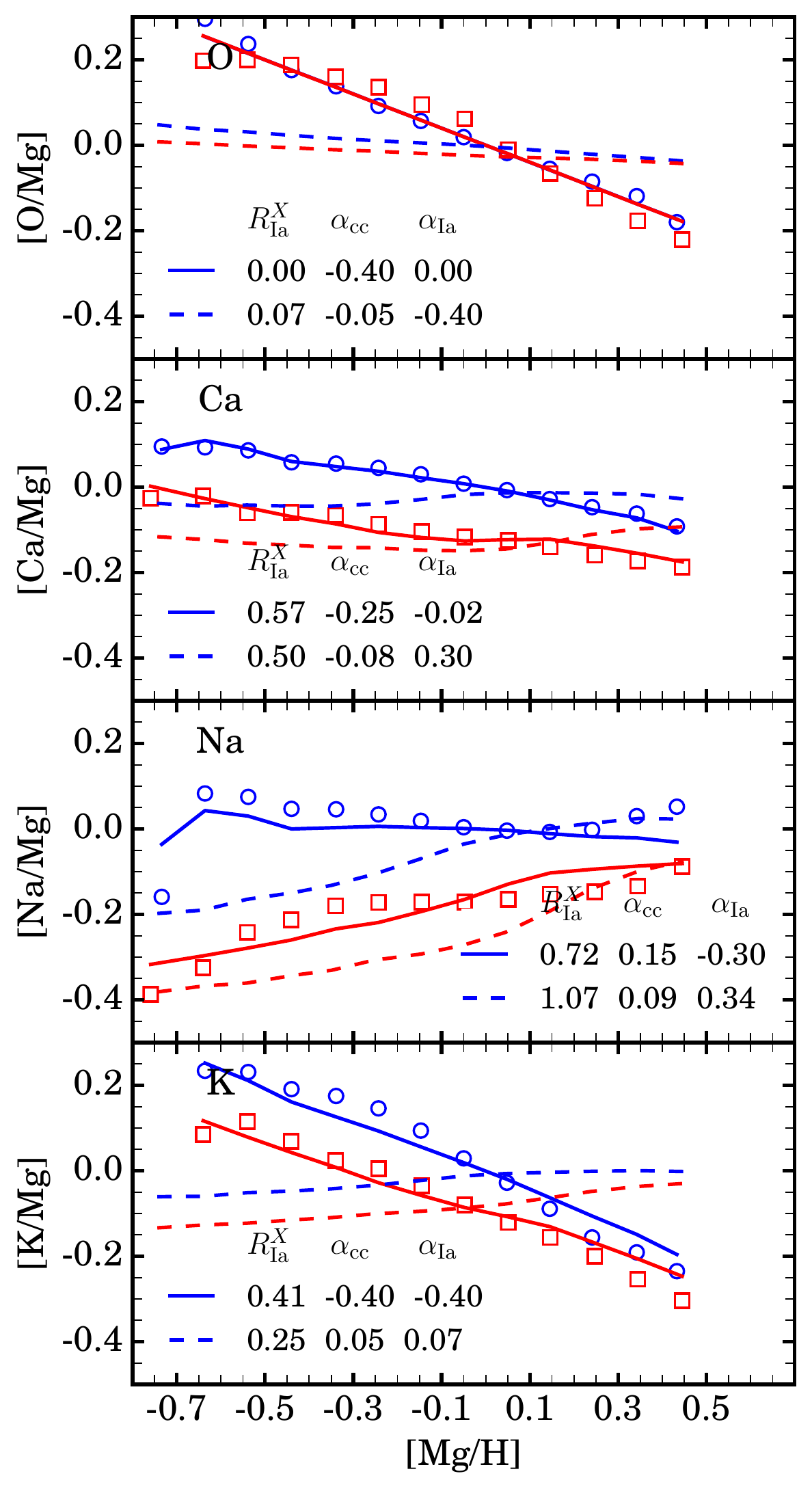}
 \caption{Similar to Figure~\ref{fig:zeros_2proc} but for remaining light-\textit{Z} elements, zero-point offsets included. We see reasonable agreement between the GALAH and APOGEE 2-process models for Ca and Na while O and K differ in $\RIa$ values and/or metallicity dependences.}
 \label{fig:lightz_2proc}
 \end{center}
\end{figure}

Figure~\ref{fig:lightz_2proc} shows the 2-process fits to the remaining light-\textit{Z} elements. Again, fit parameters for GALAH and APOGEE [X/Mg] median trends \citepalias{weinberg} are included within each plot. 

The high-Ia and low-Ia sequences for Ca are well fit with model parameters similar to those for Si: $\RIa=0.57$ and $\acc=-0.25$. APOGEE data imply a similar $\RIa$ but flatter metallicity trend. The yield models used by \citetalias{andrews} predict a larger SNIa contribution to Ca than to Si, which matches the difference found in APOGEE but not in GALAH.

While Si and Ca are $\alpha$-elements with CCSN and SNIa contribution, we expect O to be almost purely CCSN. In agreement with this theoretical prediction, the [O/Mg] 2-process fit returns an $\RIa=0.00$. For APOGEE fits we find a similarly small $\RIa$ value but a much flatter metallicity trend. As previously noted, a sloped trend of [O/Mg] vs. [Fe/H] is fairly common in optical surveys, but there is no obvious theoretical explanation for it because IMF-averaged yields are expected to have weak metallicity dependence for both O and Mg \citepalias[see, e.g.,][]{andrews}. 

The light odd-\textit{Z} element Na exhibits larger sequence separations than the $\alpha$-elements, with a best fit $\RIa = 0.72$. The APOGEE fit prefers an even larger SNIa contribution. Neither fit is an excellent match to the data, but both surveys clearly show a large [Na/Mg] separation between the high-Ia and low-Ia median sequences. This contrasts with the theoretical yield models, which predict that CCSN production of Na is much higher than either SNIa or AGB production (e.g., \citetalias{andrews}; \citealp{rybizki}).

The [K/Mg] trends in GALAH are significantly different from those in APOGEE. However, improved stellar atmosphere models in the next GALAH data release are expected to produce much flatter [K/Mg] trends (S. Buder, personal communication), so we do not put much weight on this difference. 

\subsection{Fe-peak}\label{subsec:2proc_fe}

\begin{figure*}[!htb]
 \begin{center}
 \includegraphics[width=.85\textwidth]{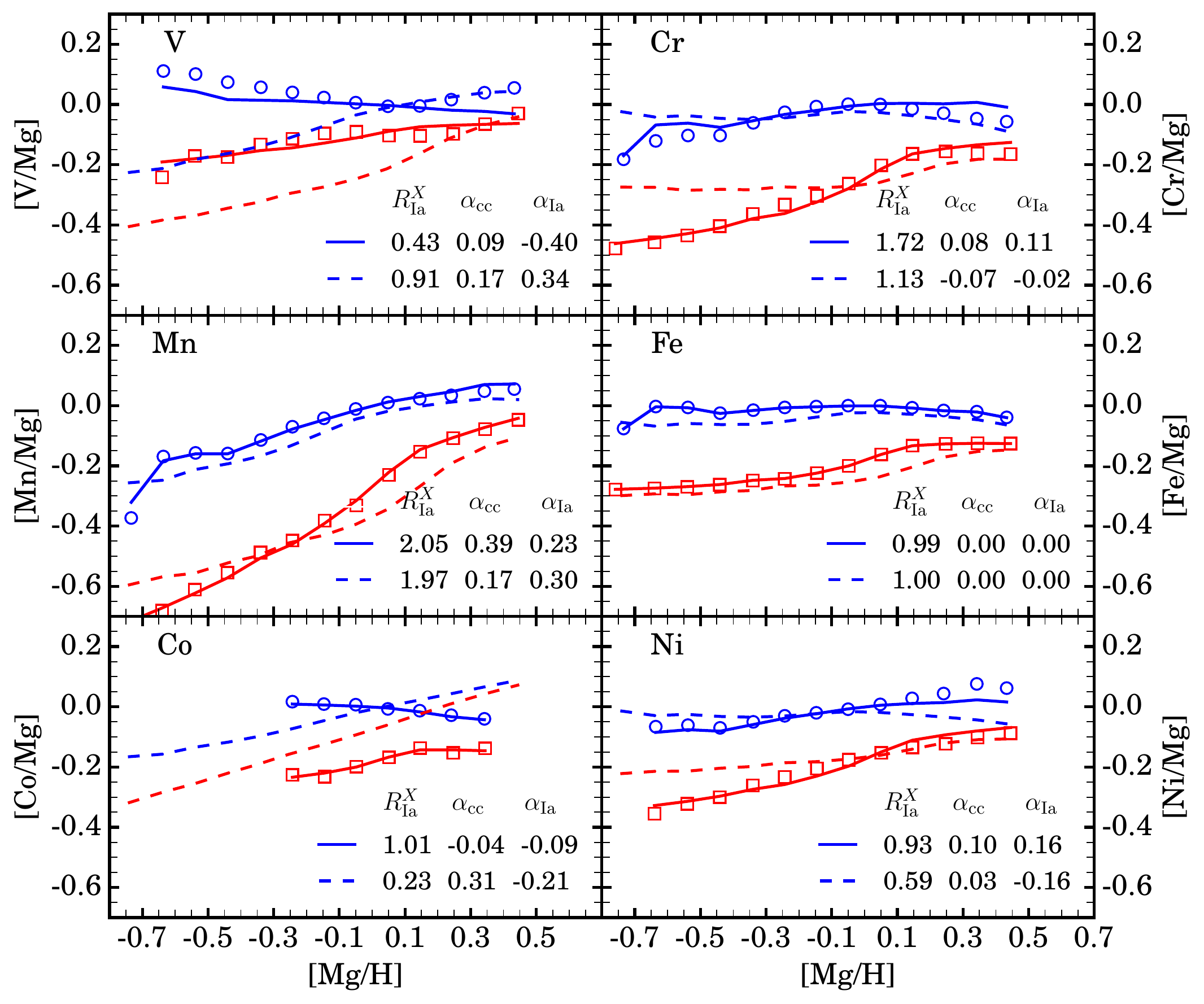}
 \caption{Same as Figure~\ref{fig:lightz_2proc} but for Fe-peak elements and with a wider y-axis. 2-process fits parameters for V and Co differ between GALAH and APOGEE while those of Mn, Cr, and Ni are similar.}
 \label{fig:fepeak_2proc}
 \end{center}
\end{figure*}

Figure~\ref{fig:fepeak_2proc} shows median sequences and 2-process model fits for Fe-peak elements. The fits for both $\femg$ sequences are perfect by construction, as the 2-process model uses $\femg$ to infer $\AIa/\Acc$, and the fit parameters are the assumed values, $\RIa=1$ and $ \acc = \aIa = 0$. One can see in Figure~\ref{fig:fepeak_2proc} that the APOGEE values of $\femg$ on the low-Ia sequence are lower, near $\mgh=0$, implying a lower SNIa contribution to these stars. As discussed previously for Si, this difference in the $\femg$ sequences leads to larger $\RIa$ values inferred from GALAH data relative to APOGEE even if the $\xmg$ sequences are similar.

All odd-\textit{Z} Fe-peak elements in APOGEE show a strong rising metallicity dependence. GALAH finds much shallower slopes, and even declining trends, for both [V/Mg] and [Co/Mg]. The 2-process fit to [V/Mg] returns $\acc=0.09$ and $\aIa = -0.40$, versus $\acc = 0.17$ and $\aIa = 0.34$ from APOGEE. The sequence separation is smaller in GALAH, implying a lower $\RIa = 0.43$ (vs. $\RIa = 0.91$ from APOGEE) despite the $\femg$ effect noted above. V is difficult to measure in GALAH, and abundances may be be affected by weak, blended lines \citepalias{buder_b}.

Co shows a roughly flat metallicity trend over the limited metallicity range available. The fitted $\acc$ and $\aIa$ are near zero, and the substantial separation of [Co/Mg] in the two sequences yields $\RIa = 1.01$, implying that about half of the Co at solar abundances comes from SNIa. Indeed, over this metallicity range the GALAH [Co/Mg] sequences are quite similar to the [Fe/Mg] sequences. The observed APOGEE trends, and thus the inferred model parameters, are quite different. 

While the trends in V and Co differ between GALAH and APOGEE, those of Mn are encouragingly similar. The strong separation between the inclined low-Ia and high-Ia sequences is fit well by the 2-process model with a large $\RIa=2.05$, similar to the $\RIa = 1.97$ for APOGEE. The observed metallicity trends differ moderately on the low-Ia sequence, leading to different separation of these trends into $\acc$ and $\aIa$; GALAH yields higher $\acc$ while APOGEE yields higher $\aIa$.  

Among the even-\textit{Z} Fe-peak elements, [Cr/Mg] shows similarities qualitatively and quantitatively to Fe. We find solar SNIa contribution of about 60$\%$ ($\RIa = 1.72$), moderately larger than inferred from APOGEE. The GALAH median trends are noticeably steeper at low metallicities than APOGEE and correspondingly are better fit by larger, positive $\acc$ and $\aIa$ values.  

Once the zero-point shifts are applied (-0.062 for Ni and +0.044 for Fe), the GALAH [Ni/Mg] and $\femg$ trends are similar. The inferred $\RIa=0.93$ is higher than for APOGEE ($\RIa=0.59$), and the GALAH data imply moderately rising metallicity trends ($\acc=0.10$, $\aIa=0.16$)

\subsection{New Fe-peak/cliff}\label{subsec:2proc_newfe}

\begin{figure}[h]
 \includegraphics[width=\columnwidth]{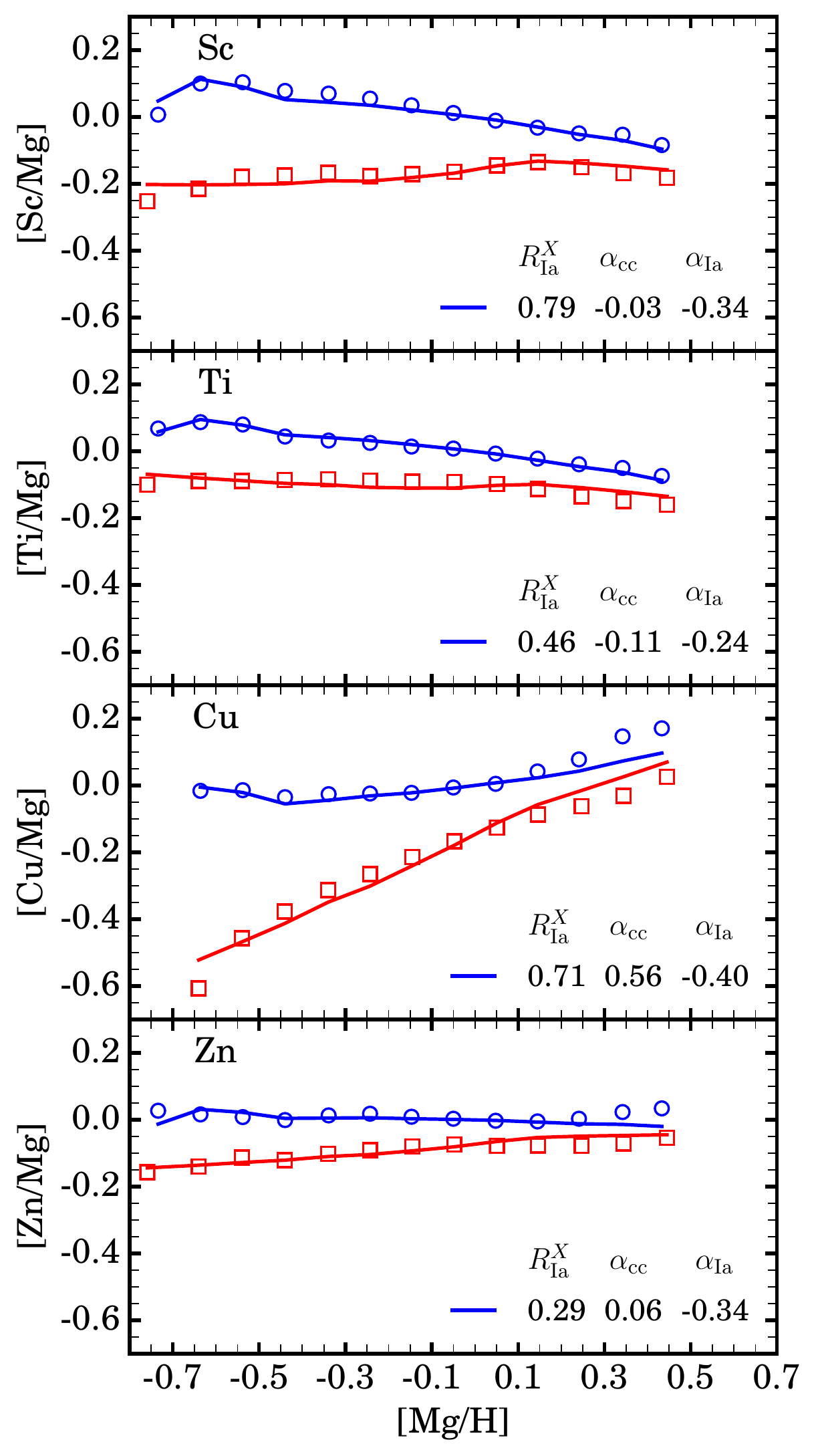}
 \caption{Same as Figure~\ref{fig:lightz_2proc} but for new Fe-peak/cliff elements and with a wider y-axis.}
 \label{fig:newsn_2proc}
\end{figure}

Next we fit the elements reported by GALAH but not by APOGEE with the 2-process model, giving insight into the production mechanisms of 9 additional elements. Figure~\ref{fig:newsn_2proc} shows median sequences and 2-process fits for elements on the ``early rise'' (Sc, Ti) and ``falling cliff'' (Cu, Zn) of the Fe-peak. The fits imply significant SNIa contributions to Sc and Ti ($\RIa=0.79$ and $0.46$, respectively). Both elements have roughly flat metallicity trends of the low-Ia population but declining trends for the high-Ia population, which leads to negative values of $\aIa$. If confirmed by future measurements, these trends could be an interesting diagnostic of SNIa explosion physics. [Zn/Mg] appears much flatter, though the 2-process model is best fit with a weak, negative SNIa metallicity dependence. Zn resembles an $\alpha$-element with a moderate SNIa contribution, such as Si. [Cu/Mg] shows a strong sequence separation, with $\RIa=0.71$. It has the largest metallicity dependence ($\acc=0.56$) of these four new elements. 

\subsection{Carbon}\label{subsec:2proc_c}

Stellar carbon is mainly produced through the Triple-Alpha reaction \citep{salpeter}. While this nuclear process is well understood, its dominant stellar sources are still debated. Early stellar models predicted that C was made in exclusively massive stars, with yields dependent upon stellar winds and rotation \citep{gustafsson, meynet}. More recent work, however, suggests that low and intermediate mass stars contribute significantly to C at higher metallicity and later times \citep[eg.][]{chiappini}. In a study of the kinematically thick and thin solar neighborhood, \citet{bensby_C} find that C traces Fe and Y, elements with substantial delayed production mechanisms from SNIa and s-process in AGB stars, respectively.  Further observational data from \citet{nissen} support C production by multiple stellar populations. Chemical evolution models have also tackled C production. They broadly conclude that low mass stars produce a substantial amount of C in the thin disk, suggesting a current contribution of 50-80\% \citep{cescutti, mattsson}.

\begin{figure}[h]
 \includegraphics[width=\columnwidth]{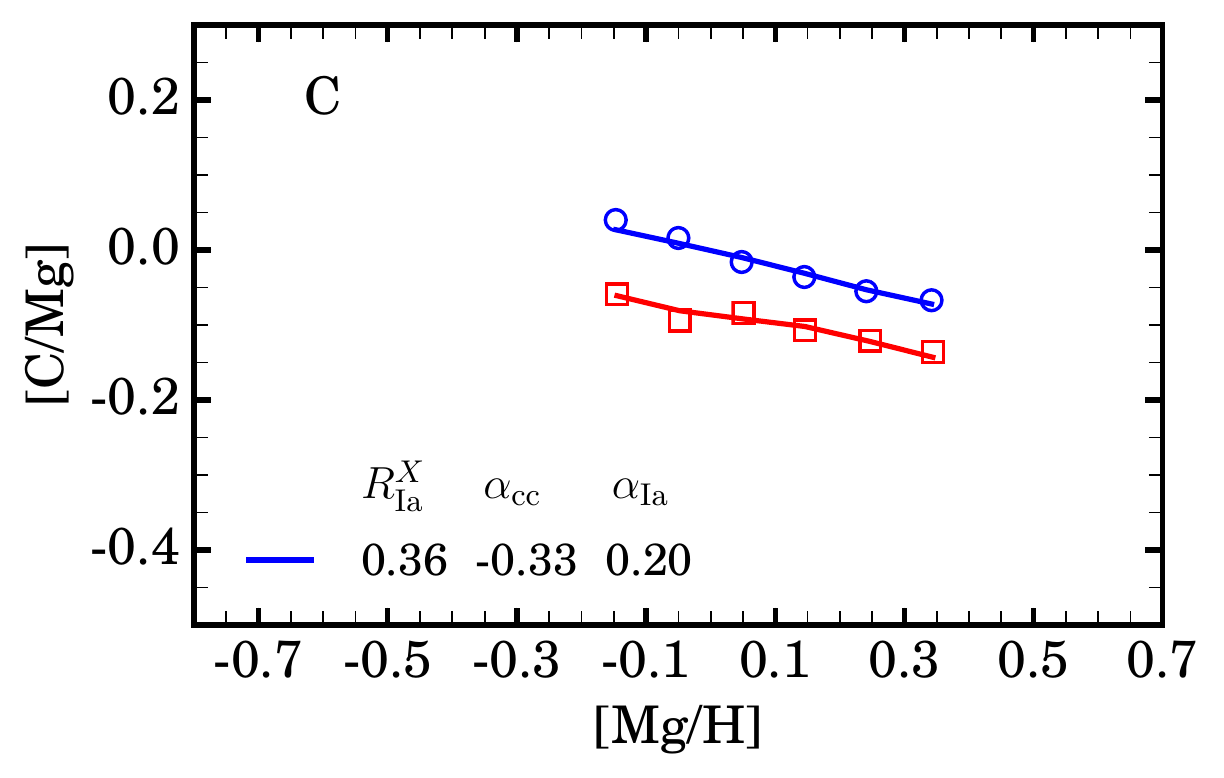}
 \caption{Same as Figure~\ref{fig:lightz_2proc} but for Carbon.  The 2-process model returns parameters suggestive of a significant delayed contribution and metallicity dependent yields.}
 \label{fig:carbon_2proc}
\end{figure}

The observational studies cited above employ small samples of stars with very high resolution spectra. We take C abundances from GALAH to form a much larger population. Measuring C in stars remains difficult, so GALAH only detects this element in a sub-sample ($\sim$ 12 000 stars) of high metallicity, hot stars where high energy atomic C lines could be found \citepalias{buder_b}. We fit the high-Ia and low-Ia sequences with the 2-process model (Figure~\ref{fig:carbon_2proc}) to constrain the delayed contribution. We find that the [C/Mg] trends are best fit with an $\RIa=0.36$, which would imply an AGB contribution of about $25\%$ if the delay time distribution matched that of SNIa. The observed trends robustly suggest both a prompt and delayed contribution to C, but quantifying the delayed fraction will require a more detailed model of the delay distribution for AGB production. The declining metallicity trends suggest declining yields toward higher metallicity, at least for the CCSN component.

\subsection{Neutron Capture}\label{subsec:2proc_neu}

\begin{figure}[h]
 \includegraphics[width=\columnwidth]{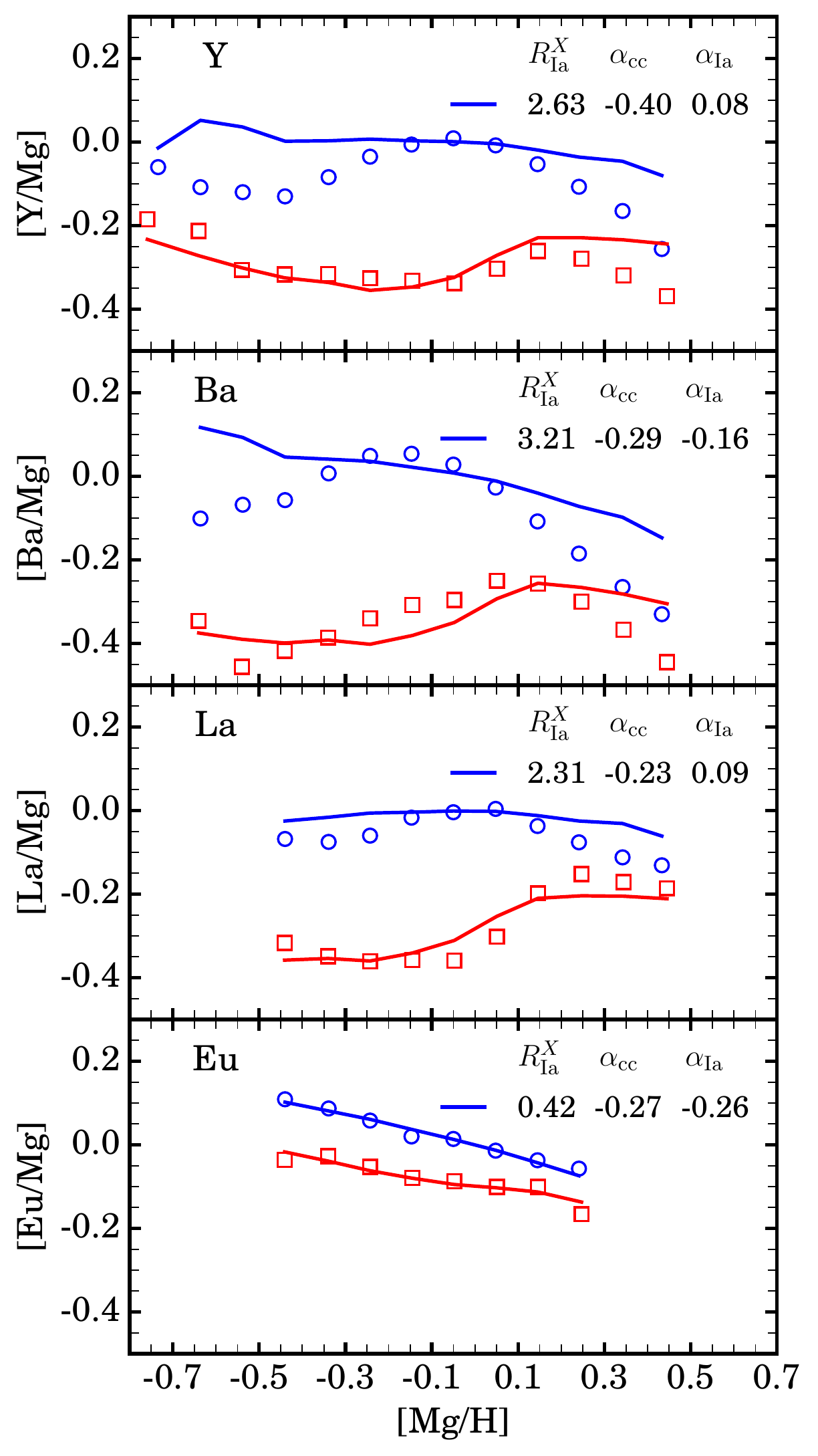}
 \caption{Same as Figure~\ref{fig:lightz_2proc} but for neutron capture elements. As expected, the 2-process model with linear metallicity dependence does not fit Y, Ba, or La well.}
 \label{fig:neutron_2proc}
\end{figure}

The previous elemental trends have been reasonably well fit by the power-law forms (linear in $\xmg$ vs. $\mgh$) incorporated in the 2-process model. The median trends of Y, Ba, and La, on the other hand, exhibit peaks near solar $\mgh$ and plateaus at low metallicity. Though we do not expect the 2-process model to describe them well, we include their fits in Figure~\ref{fig:neutron_2proc}. As expected, the model cannot reproduce the peaked sequences for [Y/Mg], [Ba/Mg], or [La/Mg]. It does give some indication to the sequence separation, returning $\RIa = 2.63$ (Y), $\RIa = 3.21$ (Ba), and $\RIa = 2.31$ (La). The sequence separation and large $\RIa$ value suggests a combination of prompt and delayed processes with more weight on the delayed component. This is in agreement with our expectation that these neutron capture elements are produced by a prompt $r$-process and a delayed $s$-process \citep{arlandini, bisterzo}.

\begin{figure}[h]
 \includegraphics[width=\columnwidth]{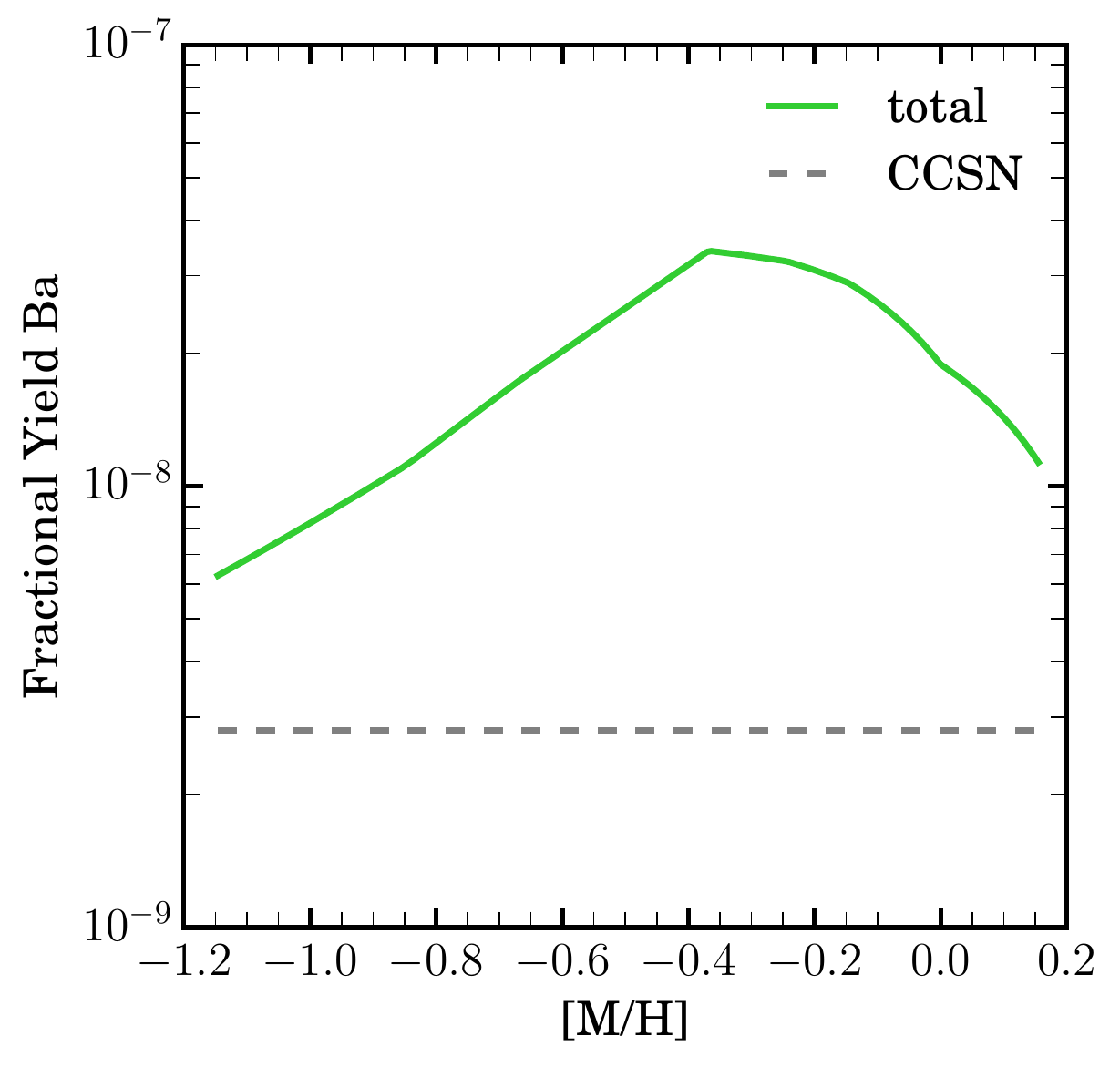}
 \caption{Fractional Ba yield from a single stellar population of mass $10^6 \text{M}_{\odot}$ and metallicity ranging from Z of 0.001 to 0.02, integrated for 10 Gyrs with VICE \citep{james}. The dashed curve denotes the CCSN component \citep{limongi18} and the solid curve denotes the total fractional yield, for which the AGB contribution \citep{cristallo} dominates.}
 \label{fig:vice}
\end{figure}

We compare our Ba metallicity trends to those of single stellar populations calculated by the Versatile Integrator for Chemical Evolution (VICE) \citep{james}. We choose CCSN yields from \citet{limongi18} and AGB yields from \citet{cristallo} and evolve a $10^6 \text{M}_{\odot}$ population for 10 Gyrs over a range of metallicities (Z=0.001 to 0.02). The resulting fractional yields are plotted vs. metallicity in Figure~\ref{fig:vice}. VICE returns low levels of fractional Ba yields from CCSN, and AGB yields which increase monotonically before turning over at [M/H] near -0.3. These fractional Ba abundances as a function of metallicity resembles GALAH's median Ba trends, without the floor at low metallicity. We conclude that the observed metallicity trends for Ba, Y, and La are in qualitative agreement with the expected metallicity dependence of AGB yields.

Unlike Y, Ba, and La, Eu is well fit by the power-law metallicity dependence of the 2-process model. The separation between the high-Ia and low-Ia sequences yields $\RIa=0.42$, evidence for two distinct Eu contributions with distinct time delay distributions. \citet{cote} have also argued for two contributions to Eu based on observed [Eu/Fe]-[Fe/H] trends. These could possibly represent a massive star and neutron star merger contribution, though the time delays for neutron star mergers are usually expected to be short compared to SNIa. The observed sequence separation and declining metallicity trends are a strong test of Eu production models. We note that the sequence separation found in GALAH is not clearly present in our analysis of the \citet{bensby_Eu} and \citet{delgado} samples.


\section{Comparison to Theoretical Yields} \label{sec:discussion}

\begin{figure*}[!htb]
 \includegraphics[width=\textwidth]{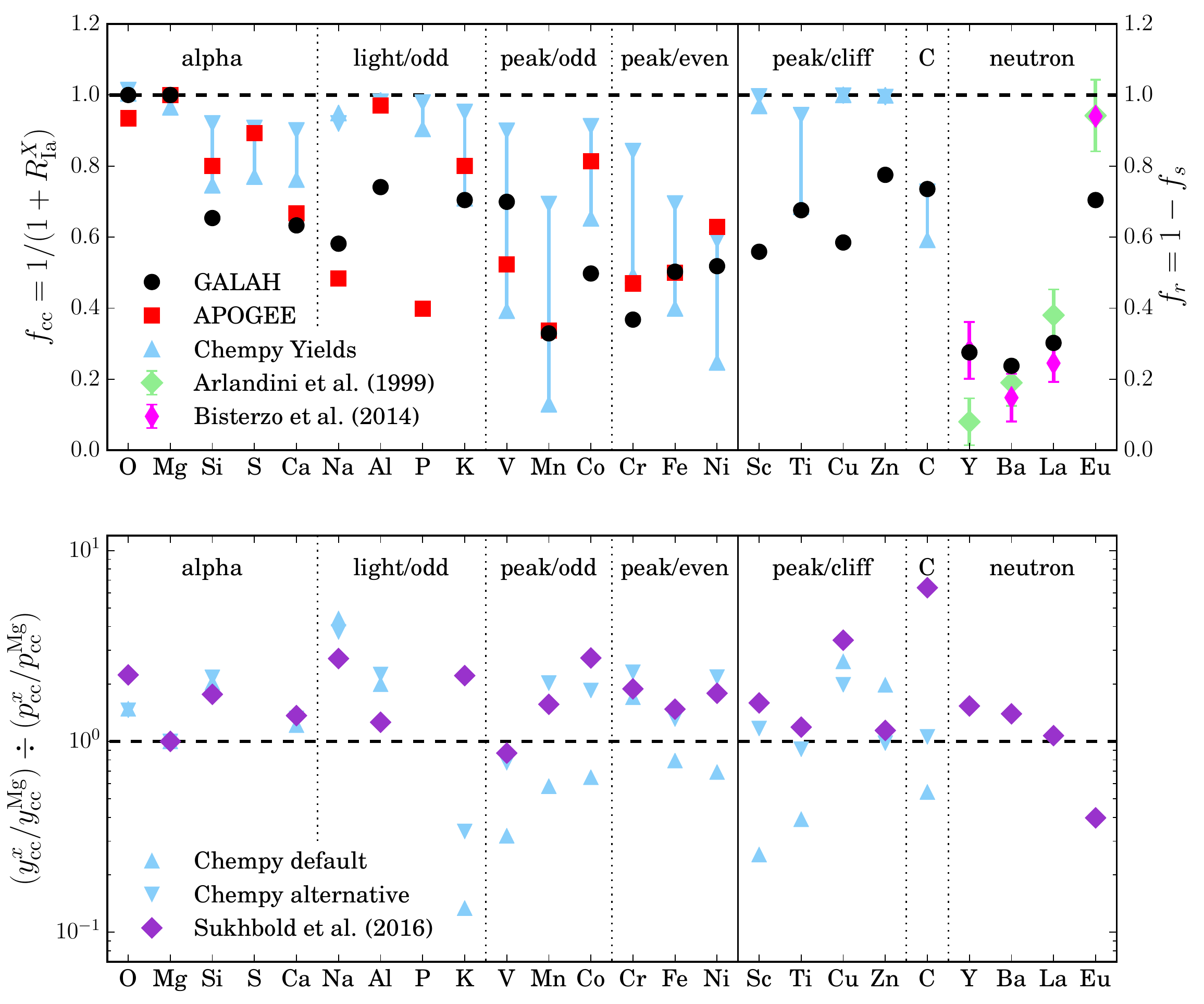}
 \caption{Top: Fractional CCSN contribution to each element at solar abundances, $\fcc=(1+\RIa)^{-1}$, inferred from the 2-process fits to GALAH median sequences (black circles). Red squares show corresponding results from APOGEE \citepalias{weinberg} for elements in common. Upward and downward pointing blue triangles show predictions of the default and alternative yields sets in \textit{Chempy}, respectively, connected to represent the range of theoretical prediction. The dashed line at $f_{cc}=1$ represents pure CCSN origin. Bottom: Ratio of theoretical CCSN yield of a given element to that of Mg, divided by the ratio of the GALAH 2-process predictions for the CCSN yield to that of Mg, all at solar abundance. Comparisons are given for both \textit{Chempy} yield sets as well as the W18 yields from \citetalias{tuguldur}. For values above/below 1, the model over/under predicts our empirically inferred CCSN yield, relative to that of Mg.
 }
 \label{fig:frac_ccsn}
\end{figure*}

To directly compare the 2-process model results with theoretical SN and neutron capture yields, we convert our ratio of SNIa to CCSN production ($\RIa$) to the fractional CCSN contribution at $\mgh = \feh = 0$: 
\begin{equation} \label{eq:fcc}
    \fcc = \frac{1}{(1+\RIa)}.
\end{equation}
We include elemental $\fcc$ values derived from the 2-process fits to GALAH data in Table~\ref{tab:zeros}. We plot these and APOGEE $\fcc$ values in the top panel of Figure~\ref{fig:frac_ccsn}, extending Figure 20 from \citetalias{weinberg}. For low-\textit{Z} and Fe-peak elements, we include yield predictions from \textit{Chempy}, a chemical evolution code from \citet{rybizki}. \textit{Chempy} returns two yield sets, a default and alternative, both plotted in Figure~\ref{fig:frac_ccsn}. The default set employs CCSN yields from \citet{nomoto}, SNIa yields from \citet{seitenzahl}, and AGB yields from \citet{karakas}. The alternative set uses CCSN, SNIa, and AGB yields from \citet{chieffi}, \citet{thielemann}, and \citet{ventura}, respectively. Both yield sets adopt a Chabrier IMF \citep{chabrier} and assume all stars between 8 and 100 $\text{M}_{\odot}$ undergo CCSN explosions. \textit{Chempy} yields presented here model the abundances of single stellar population formed at $t=0$ with solar metallicity after 10 Gyrs.

For the neutron capture elements, we include a predicted fractional $r$-process contribution, $f_r$. We assume neutron capture elements are produced solely by the $r$- and $s$-process such that 
\begin{equation}
f_r = 1-f_s,
\end{equation}
and take $s$-process contributions from \citet{arlandini} and \citet{bisterzo}. 

In the top left portion of Figure~\ref{fig:frac_ccsn}, the separation between the black circle and red square indicates the level of agreement or disagreement between GALAH and APOGEE $\fcc$ values for that element; as seen in Figures~\ref{fig:zeros_2proc}$-$\ref{fig:fepeak_2proc}, the two data sets may imply different metallicity trends even if they agree on $\fcc$ at solar metallicity. The separation between these points and the upward/downward blue triangles indicates the level of agreement or disagreement with $\textit{Chempy}$ default/alternative yield predictions, and the blue line connecting these triangles illustrates the model uncertainty associated with these different yields. The right half of this panel shows elements that are unique to GALAH. In the left half, S and P are unique to APOGEE. 

Agreement for the $\alpha$-elements is generally good. Both GALAH and APOGEE imply $\fcc\approx1$ for O (despite very different metallicity trends), in agreement with the \textit{Chempy} models. Both data sets imply $\fcc\sim0.6-0.8$ for Si and Ca, somewhat below the \textit{Chempy} predictions and closer to the default yields.

The light odd-\textit{Z} elements show larger deviations. As discussed in Section~\ref{subsec:2proc_light}, the offset between the low-Ia and high-Ia [Na/Mg] sequences seen in both GALAH and APOGEE implies a substantial non-CCSN contribution to Na, $40-50\%$ at solar abundances. However, both \textit{Chempy} yield sets predict that $\sim90\%$ of Na comes from CCSN. For Al, GALAH and APOGEE disagree substantially, and the APOGEE measurement agrees with the model predictions that nearly all Al comes from CCSN. For K, GALAH and APOGEE give similar $\fcc\sim0.75$ (but different metallicity trends), in good agreement with the \textit{Chempy} default predictions. However, both surveys struggle to measure robust K abundances, so this agreement may be partly good luck.

Moving to the Fe-peak elements, GALAH and APOGEE agree well on $\fcc$ values for Mn, Cr, and Ni (and for Fe by construction) but disagree for V and Co. Both elements are somewhat challenging for both surveys; in particular, V lines are weak and blended in GALAH spectra. For all of these elements the \textit{Chempy} default yields predict much lower $\fcc$ values (higher SNIa contributions) than the alternative yields. The data points generally lie between these two predictions, though the GALAH $\fcc$ values are below both model predictions for Co and Cr. 

For the new Fe-peak/cliff elements (Sc, Ti, Cu, Zn), the GALAH data imply $\fcc$ values ranging from 0.56 to 0.78. However, both \textit{Chempy} yield models predict $\fcc\approx1$ for Sc, Cu, and Zn. The alternative yield set also predicts $\fcc\approx1$ for Ti, but the default yield set is consistent with our inference from GALAH. As shown in \citet{rybizki}, the \textit{Chempy} models do not accurately reproduce the protosolar Ti abundance and, although they are not shown in the paper, the same is true for Sc, Cu, and Zn (J. Rybizki, personal communication). The \citet{karakas} and \citet{ventura} AGB yields used by \citet{rybizki} do not include these four elements, which might account for the models' high $\fcc$ values and low protosolar abundance predictions, or the discrepancy could arise from underpredicting SNIa yields. Similarly, \citet{tuguldur} find that CCSN yields alone (which include the weak \textit{s}-process production in the pre-supernova star) cannot reproduce solar levels of Ti, Sc, and Zn.

The alternative yield \textit{Chempy} model accurately reproduces our inferred $\fcc=0.74$ for C, while the default yield model underpredicts $\fcc$. In both models the non-CCSN contribution to C comes from AGB stars rather than SNIa, so our inferred $\fcc$ value may be inaccurate. Nonetheless, both the data and the models support the view that a substantial fraction of solar C originates in CCSN and a substantial fraction comes from another source.

For the neutron capture elements we again caution that our inferred $\fcc$ values are only qualitatively informative because the time distribution of delayed sources probably does not match that of SNIa and the prompt component is not guaranteed to be conventional CCSN. It is nonetheless encouraging that the $\fcc$ values of $\approx 0.2-0.3$ inferred for Y, Ba, and La agree reasonably well with the fraction of these elements inferred to come from the \textit{r}-process by \citet{arlandini} and \citet{bisterzo}.

Eu is usually regarded as a nearly pure \textit{r}-process element, but the separation of [Eu/Mg] in GALAH data implies that some Eu enrichment occurs with a significant time delay relative to star formation ($\fcc=0.70$). As emphasized by \citet{schonrich} who focused on abundance ratios in lower metallicity stars, some of this time delay could be associated with injection into the hot ISM phase rather than genuinely delayed elemental production. Nonetheless, the gap between the low-Ia and high-Ia [Eu/Mg] sequences suggests that there may be two distinct contributions to Eu enrichment, perhaps one associated with massive star collapse explosions \citep{metzger} and one with neutron star mergers \citep{smartt}. Neutron star mergers with a short minimum delay time but a power-law delay time distribution \citep{hotokezaka} might effectively mimic a prompt and delayed combination. 

Because our 2-process decomposition isolates the CCSN contribution to each element, we can directly test predictions for the IMF-averaged CCSN yields independent of theoretical models of SNIa and AGB yields. To do so, in the lower panel of Figure~\ref{fig:frac_ccsn} we divide the ratio ($y_{\rm cc}^{\rm X}/y_{\rm cc}^{\rm Mg}$) of the predicted CCSN yield to the predicted Mg yield by ($p_{\rm cc}^{\rm X}/p_{\rm cc}^{\rm Mg}$), the Mg-normalized value of our inferred core collapse process amplitude for that element. We use solar metallicity CCSN yields and compare to the 2-process amplitudes at $\mgfe=0$. If $(y_{\rm cc}^{\rm X}/y_{\rm cc}^{\rm Mg})\div (p_{\rm cc}^{\rm X}/p_{\rm cc}^{\rm Mg})$ for element X equals 1, then the yield model reproduces our results from the 2-process fit to the corresponding GALAH abundance. If $(y_{\rm cc}^{\rm X}/y_{\rm cc}^{\rm Mg})\div (p_{\rm cc}^{\rm X}/p_{\rm cc}^{\rm Mg})$ is greater/less than 1, then the model over/under predicts the CCSN contribution to the elemental abundance once it is normalized to produce solar (Mg/H). 

As in \citetalias{weinberg}, we compare the 2-process amplitudes to the \textit{Chempy} default and alternative yield sets, which employ grids based CCSN yields from \citet{nomoto} and \citet{chieffi}, respectively, adopting a Chabrier IMF with all stars 8-100 M$_{\odot}$ producing CCSN. We add the CCSN yields from \citet[][hereafter S16]{tuguldur} to our comparison, specifically those that adopt the W18 explosion engine. The yields from \citetalias{tuguldur} have a finer mass sampling, integrate over a Salpeter IMF \citep{salpeterIMF} extending to 120 M$_{\odot}$, and implement a neutrino powered explosion model. Rather than a mass cutoff beyond which no stars explode, \citetalias{tuguldur} find islands of explodability in their high mass star models, because O- and C-burning shells migrate during the late stages of evolution and change the compactness of the pre-supernova star in a non-monotonic manner. The choice of IMF causes some differences in the predicted CCSN yields between \textit{Chempy} and \citetalias{tuguldur}, but pre-supernova evolution and the latter's inclusion of explodability has a larger impact on the predictions. The larger number of low mass stars in the Salpeter IMF reduces all yields by a constant factor, but this effect is removed by our normalization to solar Mg abundance.

Our results in the lower panel of Figure~\ref{fig:frac_ccsn} can be compared qualitatively to those in Figure 14 of \citet{rybizki} and Figure 24 of \citetalias{tuguldur}. The former compares the predictions of the full \textit{Chempy} chemical evolution models, including SNIa and AGB contributions, to proto-solar abundances. The latter compares \citetalias{tuguldur}'s IMF-averaged CCSN yields to proto-solar abundances, normalized so that O is exactly reproduced. Both \textit{Chempy} and \citetalias{tuguldur} predict super-solar O/Mg ratios, and since we normalize to Mg here, this appears as overproduction of O. Normalizing to O instead would move all \citetalias{tuguldur} (purple) points down by a factor of 2.2 and all \textit{Chempy} yields down by a factor of 1.4, constant vertical offsets in the logarithmic plot. With Mg normalization, the \textit{Chempy} and \citetalias{tuguldur} yields overpredict Si by about a factor of two and slightly overproduce Ca. We note that the \citetalias{tuguldur} plot shows [Si/Mg] slightly above zero and [Ca/Mg] slightly below zero. Both elements show up as overproduced here because \citetalias{tuguldur} compare their CCSN yields to total proto-solar abundances while we compare to just the CCSN component of proto-solar abundances using the fraction $\fcc=0.66$ and 0.64 empirically inferred from the GALAH sequences. Overall, we regard the agreement between the theoretically predicted and empirically inferred yields for the $\alpha$-elements as reasonably good, with the most significant discrepancy being the factor $\sim 1.5-2$ underprediction of Mg relative to the others. 

For Na, the \textit{Chempy} and \citetalias{tuguldur} yields overpredict our inferred yields from GALAH by factors of about 4 and 3, respectively. Although \citet{rybizki} and \citetalias{tuguldur} both show predicted [Na/Mg]$>0$, the discrepancy is sharper here because GALAH and APOGEE both imply that $40-50\%$ of Na comes from a non-CCSN source. For Al, the \textit{Chempy} yields show a factor of two overprediction while the \citetalias{tuguldur} yields show good agreement. Both \textit{Chempy} models drastically underpredict K  (as shown by \citet{rybizki}), but the S16 yields overpredict it by a factor of $\sim 2$. This difference may arise partly from progenitor structure and partly from the different treatment of explosion physics (T. Sukhbold, personal communication). All three of these elements have substantial observational uncertainties, as demonstrated by the difference between their GALAH and APOGEE trends, but with reliable measurements they could be powerful diagnostics of massive star evolution and CCSN physics. 

We see generally acceptable agreement between the Fe-peak CCSN yields from \textit{Chempy} and \citetalias{tuguldur} and those inferred from GALAH. The \textit{Chempy} default and alternative sets bracket (or nearly bracket) 1:1 agreement for most of these elements, though since the two predictions sometimes span a factor of four this is not a high bar. For the \citetalias{tuguldur} yields there are two striking disagreements: Co overpredicted by a factor of 3, and Cu overpredicted by a factor of 4. In both cases these discrepancies are larger than one would see in a simple comparison of model yields to proto-solar abundances because the GALAH sequence separation implies that roughly half of the solar Co and Cu comes from non-CCSN sources. While verification with other data is desirable (GALAH and APOGEE disagree for Co and Cu is unique to GALAH), these overpredictions could be a useful diagnostic for supernova models.

The \textit{Chempy} alternative yield set agrees well with the C yield inferred from GALAH, while the default set underpredicts the GALAH value (relative to Mg) by about a factor of two. Figure 14 of \citet{rybizki} shows similar behavior. The \citetalias{tuguldur} model, on the other hand, overpredicts C by a factor of $\sim7$. This discrepancy is larger than the factor of $\sim2$ overproduction shown in \citetalias{tuguldur} because we normalize to Mg instead of O and because we attribute $26\%$ of C to non-CCSN sources. The origin of the high predicted C abundances is not clear, but uncertainties in the triple-$\alpha$ and $^{12} \text{C} (\alpha,\gamma) ^{16}\text{O}$ reactions could be one source \citep{fields}.

For the neutron capture elements Y, Ba, and La, the \citetalias{tuguldur} yields agree well with our inferred ``prompt'' component--the component that tracks CCSN enrichment. Eu is underpredicted by a factor of $\sim2$. Given the uncertainty of our CCSN/non-CCSN decomposition for these elements, we regard this as an encouraging level of agreement, even in the case of Eu. In future work we will examine the modeling approaches that carry out this decomposition using additional sources beyond CCSN and SNIa. The clear separation in $\xmg$ values for stars with high and low $\femg$, seen in GALAH for all four of these elements, will be an essential constraint in disentangling the populations of stars that produce them. 

\section{Summary} \label{sec:summary}

Using data from GALAH DR2, we present and interpret the median trends of $\xmg$ vs. $\mgh$ for 21 elements: O, Si, Ca, Na, Al, K, V, Mn, Co, Cr, Fe, Ni, Ti, Sc, Cu, Zn, C, Y, Ba, La, and Eu. After temperature, quality, and SNR cuts, our sample comprises 70 924 stars spanning the metallicity range $-0.8\lesssim \mgh \lesssim +0.5$, though for individual elements the number of stars may be smaller and the metallicity range more restricted. Twelve of these elements were previously studied using APOGEE DR14 abundances \citep{holtzman} by \citetalias{weinberg}, and comparison of results derived by different pipelines from optical and near-IR spectra helps to show which trends are robust and which are sensitive to systematic uncertainties in the observations or the abundance modeling. GALAH provides access to several new elements including C, the intermediate elements Sc and Ti, the ``Fe-cliff'' elements Cu and Zn, and the neutron capture elements Y, Ba, La, and Eu.

Following \citetalias{weinberg}, we separately construct median sequences for stars in the low-Ia (``high-$\alpha$'') and high-Ia (``low-$\alpha$'') populations defined by a boundary in the $\mgfe-\feh$ plane. Although our GALAH sample consists primarily of main sequence stars relatively close to the Sun, \citetalias{weinberg} find that the median $\xmg-\mgh$ sequences are nearly independent of location in the Galactic disk, even though the $\mgh$ distribution and relative numbers of low-Ia and high-Ia stars depends strongly on Galactic $R$ and $|Z|$. This universality suggests that these sequences depend primarily on IMF-averaged nucleosynthetic yields with little sensitivity to other aspects of chemical evolution. The separation of $\xmg$ low-Ia and high-Ia sequences of populations indicate the fraction of element X produced by prompt CCSN vs. delayed sources that track SNIa. The slopes (or more generally the shapes) of these sequences are diagnostic of metallicity dependent yields. Because these sequences are approximately universal (at least in APOGEE) and the intrinsic scatter about them is small, comparing them across surveys is a stringent test for consistency of abundance scales. Comparisons of full-population trends in [X/Fe] vs. $\feh$ can mask discrepancies because they have larger scatter and are affected by sample selection (e.g, by the relative proportion of thin and thick disk stars). 

We fit the GALAH sequences using the 2-process model of \citetalias{weinberg}, in which the abundances of any star are the sum of a CCSN process and an SNIa process, and the ratio (X/Fe) within either process has a power law dependence on (Mg/H). More generally one can regard the 2-process fit as decomposing the contributions to a given element into a prompt component that tracks CCSN enrichment and a delayed component that tracks SNIa enrichment. A power-law metallicity trend appears adequate for many of the elements we consider over our limited range of $\mgh$, but it is clearly a poor approximation for Y, Ba, and La. The 2-process model assumes that all Mg comes from CCSN and that the IMF-averaged yields of Mg and Fe are independent of metallicity. Before fitting models, we apply zero-point offsets (typically $\sim0.08$ dex, see Table~\ref{tab:zeros}) to the GALAH abundances so that the high-Ia sequences pass through $\xmg=0$ at $\mgh=0$.

We observe the following:
\begin{enumerate}
    \item GALAH and APOGEE agree fairly well in the $\xmg$ sequence separation and slopes for Si, Mn, Cr, Fe, and Ni. For other elements the sequence separation (Al, V, Co) and/or metallicity trends (Al, O, Ca, Na, K, V, Co) are substantially different.
    \item Both surveys imply a nearly pure-CCSN origin for O (i.e., no sequence separation) and a significant but sub-dominant SNIa contribution to Si and Ca.
    \item Like other optical surveys, GALAH shows an [O/Mg] trend that rises towards low metallicities, while APOGEE shows a flag [O/Mg] trend. The APOGEE trend agrees better with theoretical yield models, which predict that O and Mg are produced by similar stars with little metallicity dependence.
    \item GALAH trends imply a moderate non-CCSN contribution to Al with little overall metallicity dependence while APOGEE trends imply a pure-CCSN origin with a weakly increasing metallicity dependent yield.
    \item Both surveys show a significant separation of [Na/Mg] sequences implying a substantial contribution to Na from non-CCSN sources. This finding conflicts with theoretical yield models.
    \item Both surveys find a large SNIa contribution to Mn, the largest of any Fe-peak element, with implied yields that increase rapidly with metallicity. 
    \item For C, GALAH shows a significant sequence separation implying a significant but sub-dominant non-CCSN contribution, and [C/Mg] trends that decrease with increasing metallicity over the observed range $-0.2\lesssim \mgh \lesssim +0.4$.
    \item GALAH trends imply significant SNIa contributions to Sc, Ti, and Cu and a small SNIa contribution to Zn. The [Cu/Mg] sequences for low-Ia stars rises steeply with $\mgh$, while the other trends for these elements are roughly flat.
    \item GALAH sequences for [Y/Mg], [Ba/Mg], and [La/Mg] show large separations for low-Ia and high-Ia populations, implying a large contribution to these elements from a delayed source, presumably AGB stars. The high-Ia sequences for all three elements show clear maxima at $\feh\approx-0.1$, in agreement with AGB yield models that predict a transition from rising to falling metallicity dependence. The trend maxima for the low-Ia and high-Ia populations occur at similar $\feh$ but offset $\mgh$, which suggests that Fe-peak nuclei are the seeds for formation of these heavier elements. 
    \item GALAH sequences for [Eu/Mg] show a clear separation for the low-Ia and high-Ia populations, implying that some Eu enrichment is substantially delayed relative to star formation. This surprising result is not clearly evident in the smaller \citet{bensby_Eu} and \citet{delgado} samples, and it merits future investigation. 
\end{enumerate}

Figure~\ref{fig:frac_ccsn} presents a summary comparison between our inferences from the 2-process model fits at solar abundances ($\mgh=\femg=0$) and the yield predictions of the default and alternative \textit{Chempy} models \citep{rybizki} and \citetalias{tuguldur} CCSN models. For most elements the inferred core collapse fraction ($\fcc$) is bracketed by or reasonably close to the \textit{Chempy} model predictions. The low inferred $\fcc$ values for Na and Al are clear exceptions, as the \textit{Chempy} models predict that CCSN should dominate the production of both elements. We also infer substantial SNIa contributions for Sc, Cu, and Zn, while the \textit{Chempy} models predict almost pure CCSN production. Our inferred $\fcc=0.76$ for C agrees well with the \textit{Chempy} models, in which non-CCSN C production comes from AGB stars. Our low inferred $\fcc\approx0.2-0.3$ values for Y, Ba, and La agree reasonably well with the $r$-process fractions found by \citet{arlandini} and \citet{bisterzo}.

All three CCSN yield models underpredict Mg production relative to O and Si, as found previously by \citetalias{tuguldur} and \citet{rybizki}. If the predicted Mg yields were boosted by a factor of $\sim1.5-2$, then the \textit{Chempy} CCSN yield models would bracket or be close to our empirical inferences for most elements, and the \citetalias{tuguldur} yields would be in reasonable agreement for most elements. The \textit{Chempy} models drastically underpredict K production, and all three models overpredict the CCSN contribution to Na. The most striking disagreements for \citetalias{tuguldur} are Cu and C, which are overpredicted by factors of $\sim4$ and $\sim7$ relative to Mg (discrepancies relative to O are a factor of 2 smaller). While further observational investigation is desirable for both elements, these discrepancies could prove to be valuable diagnostics for massive star evolution or supernova explosion physics. The \citetalias{tuguldur} yields for Y, Ba, and La agree well with our inferred prompt contribution to these elements, while the Eu yield can account for roughly half of our inferred prompt component.

The approaches to multi-element abundance analysis used in \citetalias{weinberg} and here offer a promising route to deriving empirical constraints on IMF-averaged nucleosynthetic yields. These empirical yields can provide valuable tests of stellar evolution and supernova models, and they can mitigate the sensitivity of chemical evolution predictions to theoretical yield uncertainties. On the modeling side, the obvious direction for future work is to extend the 2-process model to include AGB enrichment and other channels that could be important for some elements. Applications to multi-zone chemical evolution models can test the accuracy of the 2-process or multi-process model in deriving nucleosynthesis constraints that are insensitive to assumptions about star formation history and gas flows.

The other clear direction for future work is improved abundance measurements. The difference between GALAH and APOGEE sequences for several of their common elements shows that systematic errors in the measured abundances remain a serious problem. Data-driven cross-calibration of surveys \citep[e.g.,][]{ness} should improve consistency of results, but we caution that getting consistent abundance trends from different surveys does not guarantee that these trends are correct. Continued improvement of stellar atmosphere models is essential, including 3D and non-LTE effects and observational tests that demonstrate good agreement between predicted and observed atmospheric structure. Alongside these improvements, extension of GALAH, APOGEE, and other multi-element surveys to other regions of the Milky Way and nearby galaxies will test whether yields from stellar nucleosynthesis are essentially universal or instead vary with time or environment.

\section*{Acknowledgements}

We thank  Sven Buder, James Johnson, Karin Lind, Jan Rybizki, and Tuguldur Sukhbold for helpful conversations and comments on the manuscript and for providing model results shown in Figures~\ref{fig:vice} and~\ref{fig:frac_ccsn}.

The GALAH survey is based on observations made at the Australian Astronomical Observatory, under programmes A/2013B/13, A/2014A/25, A/2015A/19, A/2017A/18. We acknowledge the traditional owners of the land on which the AAT stands, the Gamilaraay people, and pay our respects to elders past and present.

This paper includes data that has been provided by AAO Data Central (datacentral.aao.gov.au).

This research made use of Astropy,\footnote{http://www.astropy.org} a community-developed core Python package for Astronomy \citep{astropy:2013, astropy:2018}.

\software{\textit{Chempy} \citep{rybizki}, Astropy \citep{astropy:2013, astropy:2018}}

\begin{figure*}[h]
\begin{center}
 \includegraphics[width=1.2\textwidth, angle=90]{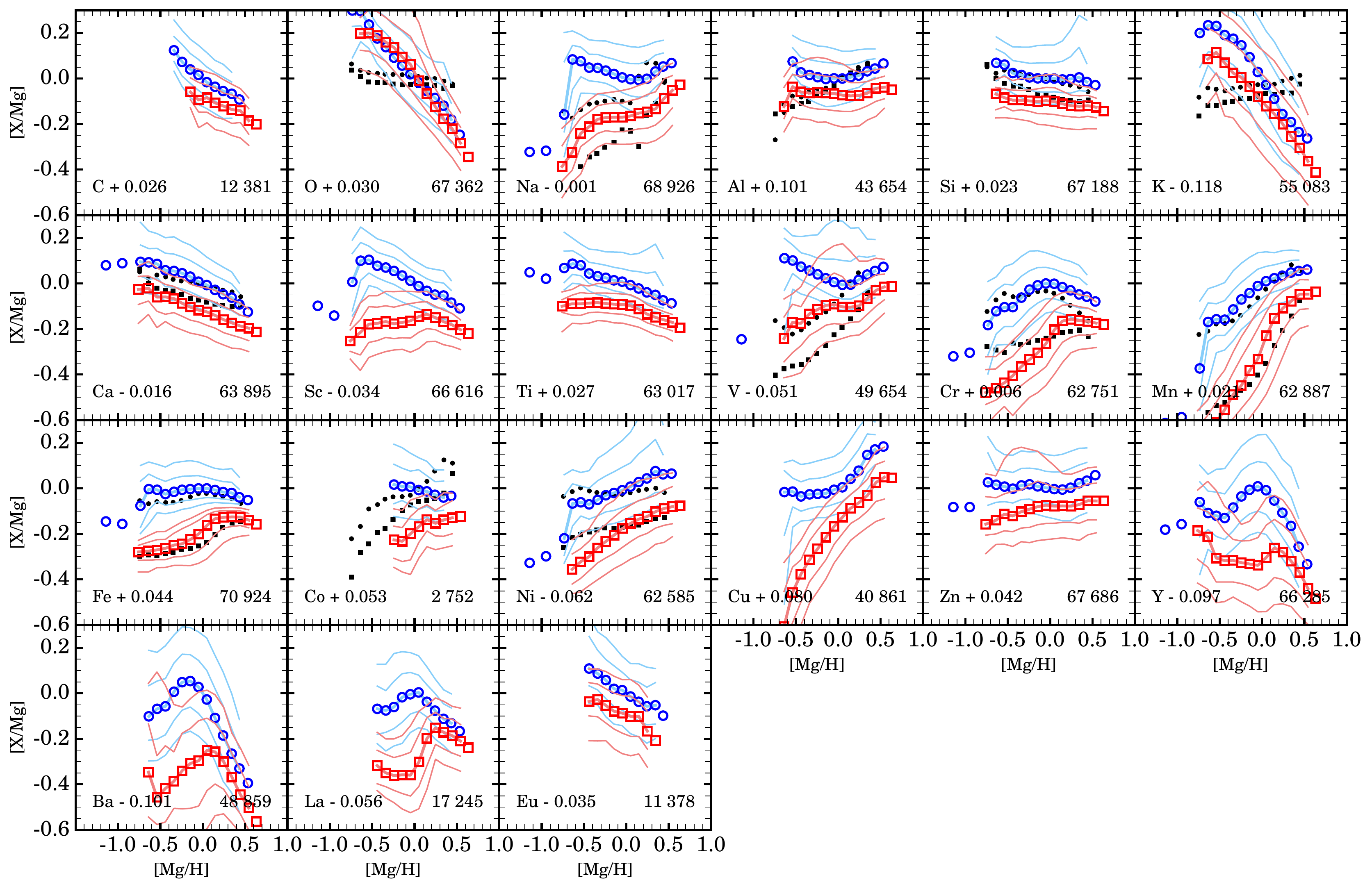}
 \caption{GALAH median abundances of the high-Ia (blue circles) and low-Ia (red squares) populations with contours at the 10th, 25th, 50th, 75th, and 90th percentiles. Data were binned by 0.1 dex in [Mg/H] space. Medians are shown for bins with $>$40 data points. APOGEE median abundances from \citetalias{weinberg} (black markers) are included where applicable. The number of stars included in the elemental population which pass our data cuts is included in the bottom right corner. In contrast to Figures~\ref{fig:alpha_med}-~\ref{fig:neutron_med}, we have applied the zero-point offsets of Table~\ref{tab:zeros} (listed in the bottom left hand corner), which effectively forces $\xmg=0$ at $\mgh=0$ along the high-Ia sequence for both GALAH and APOGEE.}
 \label{fig:all_meds}
 \end{center}
\end{figure*}

\appendix
We include tables of all GALAH median abundance trends (Tables~\ref{tab:overlap_meds}-~\ref{tab:new_meds}), ordered by elemental number, and a plot of all median sequences (Figure ~\ref{fig:all_meds}) for reference. All median values and plots include the zero-point offsets.

\begin{deluxetable*}{ccccccccccccc}[h]
\tablecaption{Median high-Ia (top) and low-Ia (bottom) sequences \label{tab:overlap_meds}}
\tablehead{
\colhead{[Mg/H]} & \colhead{[O/Mg]} & \colhead{[Si/Mg]} & \colhead{[Ca/Mg]} & \colhead{[Na/Mg]} & \colhead{[Al/Mg]} & \colhead{[K/Mg]} & \colhead{[V/Mg]} & \colhead{[Mn/Mg]} & \colhead{[Co/Mg]} & \colhead{[Cr/Mg]} & \colhead{[Fe/Mg]} & \colhead{[Ni/Mg]}
}
\startdata
-1.142 & -- & -- & 0.080 & -0.322 & -- & -- & -0.245 & -0.613 & -- & -0.320 & -0.145 & -0.327 \\ 
-0.949 & -- & -- & 0.089 & -0.317 & -- & -- & -- & -0.587 & -- & -0.304 & -0.156 & -0.298 \\ 
-0.734 & 0.299 & -- & 0.095 & -0.158 & -- & 0.200 & -- & -0.373 & -- & -0.183 & -0.076 & -0.219 \\ 
-0.636 & 0.296 & 0.069 & 0.093 & 0.084 & -- & 0.234 & 0.111 & -0.169 & -- & -0.122 & -0.003 & -0.066 \\ 
-0.538 & 0.237 & 0.061 & 0.086 & 0.076 & 0.075 & 0.231 & 0.101 & -0.157 & -- & -0.104 & -0.006 & -0.061 \\ 
-0.44 & 0.176 & 0.024 & 0.058 & 0.048 & 0.027 & 0.191 & 0.074 & -0.159 & -- & -0.104 & -0.025 & -0.070 \\ 
-0.339 & 0.138 & 0.016 & 0.055 & 0.047 & 0.016 & 0.175 & 0.057 & -0.114 & -- & -0.062 & -0.015 & -0.050 \\ 
-0.243 & 0.092 & 0.005 & 0.045 & 0.035 & 0.006 & 0.146 & 0.040 & -0.070 & -0.052 & -0.027 & -0.006 & -0.030 \\ 
-0.147 & 0.057 & 0.001 & 0.030 & 0.020 & -0.002 & 0.094 & 0.023 & -0.042 & -0.060 & -0.008 & -0.003 & -0.020 \\ 
-0.05 & 0.019 & 0.000 & 0.008 & 0.005 & 0.001 & 0.029 & 0.006 & -0.011 & -0.062 & 0.000 & 0.000 & -0.008 \\ 
0.048 & -0.018 & 0.000 & -0.007 & -0.003 & 0.000 & -0.028 & -0.006 & 0.010 & -0.076 & -0.001 & 0.000 & 0.008 \\ 
0.145 & -0.055 & -0.004 & -0.028 & -0.006 & 0.006 & -0.089 & -0.005 & 0.023 & -0.082 & -0.016 & -0.007 & 0.028 \\
0.241 & -0.085 & -0.002 & -0.047 & -0.001 & 0.013 & -0.156 & 0.016 & 0.034 & -0.097 & -0.030 & -0.016 & 0.044 \\ 
0.342 & -0.119 & 0.005 & -0.062 & 0.031 & 0.032 & -0.191 & 0.039 & 0.048 & -0.109 & -0.048 & -0.020 & 0.076 \\ 
0.433 & -0.180 & -0.015 & -0.092 & 0.053 & 0.044 & -0.235 & 0.055 & 0.055 & -0.102 & -0.058 & -0.039 & 0.062 \\ 
0.533 & -0.247 & -0.029 & -0.125 & 0.068 & 0.066 & -0.263 & 0.073 & 0.061 & -- & -0.079 & -0.051 & 0.065 \\
\hline
-0.759 & -- & -- & -0.026 & -0.386 & -- & -- & -- & -0.747 & -- & -0.479 & -0.279 & -- \\ 
-0.64 & 0.198 & -0.068 & -0.021 & -0.324 & -0.123 & 0.085 & -0.242 & -0.681 & -- & -0.459 & -0.275 & -0.355 \\ 
-0.54 & 0.200 & -0.084 & -0.060 & -0.241 & -0.036 & 0.115 & -0.171 & -0.611 & -- & -0.436 & -0.270 & -0.322 \\ 
-0.441 & 0.188 & -0.095 & -0.059 & -0.212 & -0.059 & 0.069 & -0.175 & -0.555 & -- & -0.405 & -0.263 & -0.300 \\ 
-0.34 & 0.160 & -0.095 & -0.066 & -0.179 & -0.064 & 0.024 & -0.133 & -0.488 & -- & -0.365 & -0.249 & -0.261 \\ 
-0.243 & 0.136 & -0.098 & -0.087 & -0.171 & -0.062 & 0.005 & -0.114 & -0.448 & -0.295 & -0.334 & -0.243 & -0.233 \\ 
-0.145 & 0.095 & -0.101 & -0.104 & -0.170 & -0.063 & -0.035 & -0.096 & -0.382 & -0.301 & -0.304 & -0.224 & -0.205 \\ 
-0.048 & 0.062 & -0.097 & -0.117 & -0.170 & -0.066 & -0.080 & -0.091 & -0.331 & -0.268 & -0.264 & -0.200 & -0.176 \\ 
0.051 & -0.010 & -0.101 & -0.125 & -0.164 & -0.073 & -0.121 & -0.103 & -0.230 & -0.237 & -0.203 & -0.162 & -0.153 \\ 
0.146 & -0.066 & -0.111 & -0.140 & -0.152 & -0.076 & -0.156 & -0.104 & -0.153 & -0.206 & -0.165 & -0.133 & -0.135 \\ 
0.247 & -0.124 & -0.119 & -0.159 & -0.146 & -0.074 & -0.200 & -0.098 & -0.108 & -0.222 & -0.157 & -0.127 & -0.123 \\ 
0.344 & -0.177 & -0.122 & -0.173 & -0.133 & -0.064 & -0.254 & -0.065 & -0.078 & -0.207 & -0.163 & -0.125 & -0.102 \\ 
0.445 & -0.221 & -0.119 & -0.187 & -0.087 & -0.043 & -0.304 & -0.030 & -0.047 & -0.198 & -0.166 & -0.126 & -0.088 \\ 
0.543 & -0.284 & -0.126 & -0.196 & -0.052 & -0.038 & -0.363 & -0.015 & -0.047 & -0.192 & -0.174 & -0.139 & -0.080 \\ 
0.633 & -0.345 & -0.142 & -0.213 & -0.028 & -0.049 & -0.412 & -0.013 & -0.036 & -- & -0.181 & -0.157 & -0.077 \\ 
\enddata
\tablecomments{Median abundances for the high-Ia (top) and low-Ia (bottom) sequences for elements in GALAH that overlap with APOGEE. Medians are calculated in bins of width 0.1 dex in $\mgh$ with $>40$ stars after implementing quality cuts discussed in Section~\ref{sec:methods}. Zero-point shifts have been included}
\end{deluxetable*}

\begin{deluxetable*}{cccccccccc}[h]
\tablecaption{Median high-Ia (top) and low-Ia (bottom) sequences \label{tab:new_meds}}
\tablehead{
\colhead{[Mg/H]} & \colhead{[Ti/Mg]} & \colhead{[Sc/Mg]} & \colhead{[Cu/Mg]} & \colhead{[Zn/Mg]} & \colhead{[C/Mg]} & \colhead{[Y/Mg]} & \colhead{[Ba/Mg]} & \colhead{[La/Mg]} & \colhead{[Eu/Mg]}
}
\startdata
-1.142 & 0.049 & -0.098 & -- & -0.082 & -- & -0.181 & -- & -- & -- \\ 
-0.949 & 0.021 & -0.141 & -- & -0.082 & -- & -0.157 & -- & -- & -- \\ 
-0.734 & 0.068 & 0.007 & -- & 0.027 & -- & -0.060 & -- & -- & -- \\ 
-0.636 & 0.087 & 0.100 & -0.016 & 0.016 & -- & -0.108 & -0.101 & -- & -- \\ 
-0.538 & 0.080 & 0.104 & -0.014 & 0.008 & -- & -0.120 & -0.068 & -- & -- \\ 
-0.44 & 0.044 & 0.078 & -0.035 & -0.001 & -- & -0.130 & -0.057 & -0.068 & 0.109 \\ 
-0.339 & 0.032 & 0.070 & -0.026 & 0.013 & 0.124 & -0.084 & 0.007 & -0.075 & 0.087 \\ 
-0.243 & 0.025 & 0.055 & -0.024 & 0.018 & 0.073 & -0.035 & 0.049 & -0.060 & 0.058 \\ 
-0.147 & 0.014 & 0.035 & -0.022 & 0.009 & 0.040 & -0.006 & 0.054 & -0.017 & 0.020 \\ 
-0.05 & 0.008 & 0.012 & -0.006 & 0.003 & 0.016 & 0.009 & 0.028 & -0.004 & 0.014 \\ 
0.048 & -0.007 & -0.011 & 0.005 & -0.003 & -0.016 & -0.008 & -0.027 & 0.004 & -0.014 \\ 
0.145 & -0.022 & -0.032 & 0.042 & -0.005 & -0.036 & -0.053 & -0.108 & -0.037 & -0.037 \\ 
0.241 & -0.039 & -0.049 & 0.078 & 0.003 & -0.055 & -0.107 & -0.185 & -0.076 & -0.057 \\ 
0.342 & -0.050 & -0.053 & 0.147 & 0.023 & -0.067 & -0.165 & -0.265 & -0.112 & -0.052 \\ 
0.433 & -0.074 & -0.084 & 0.171 & 0.034 & -0.092 & -0.256 & -0.330 & -0.131 & -0.098 \\ 
0.533 & -0.087 & -0.109 & 0.184 & 0.058 & -- & -0.333 & -0.394 & -0.166 & -- \\ 
\hline
-0.759 & -0.100 & -0.252 & -- & -0.157 & -- & -0.185 & -- & -- & -- \\ 
-0.64 & -0.089 & -0.215 & -0.607 & -0.140 & -- & -0.213 & -0.346 & -- & -- \\ 
-0.54 & -0.089 & -0.179 & -0.457 & -0.113 & -- & -0.306 & -0.456 & -- & -- \\ 
-0.441 & -0.086 & -0.175 & -0.377 & -0.121 & -- & -0.317 & -0.418 & -0.317 & -0.036 \\ 
-0.34 & -0.084 & -0.167 & -0.313 & -0.102 & -- & -0.316 & -0.386 & -0.349 & -0.027 \\ 
-0.243 & -0.088 & -0.177 & -0.265 & -0.091 & -- & -0.326 & -0.340 & -0.361 & -0.053 \\ 
-0.145 & -0.091 & -0.171 & -0.214 & -0.080 & -0.059 & -0.332 & -0.308 & -0.358 & -0.079 \\ 
-0.048 & -0.092 & -0.164 & -0.167 & -0.074 & -0.094 & -0.338 & -0.296 & -0.359 & -0.087 \\ 
0.051 & -0.098 & -0.145 & -0.126 & -0.078 & -0.084 & -0.303 & -0.250 & -0.302 & -0.101 \\ 
0.146 & -0.113 & -0.135 & -0.087 & -0.077 & -0.107 & -0.261 & -0.257 & -0.198 & -0.101 \\ 
0.247 & -0.135 & -0.150 & -0.062 & -0.078 & -0.121 & -0.279 & -0.300 & -0.152 & -0.166 \\ 
0.344 & -0.149 & -0.168 & -0.031 & -0.071 & -0.136 & -0.319 & -0.367 & -0.171 & -0.208 \\ 
0.445 & -0.160 & -0.182 & 0.026 & -0.054 & -0.141 & -0.369 & -0.445 & -0.186 & -- \\ 
0.543 & -0.171 & -0.203 & 0.049 & -0.055 & -0.185 & -0.440 & -0.503 & -0.210 & -- \\ 
0.633 & -0.195 & -0.220 & 0.046 & -0.055 & -0.200 & -0.484 & -0.563 & -0.239 & -- \\ 
\enddata
\tablecomments{Median abundances for the high-Ia (top) and low-Ia (bottom) sequences for elements in GALAH that are not included in APOGEE. Medians are calculated in bins of width 0.1 dex in $\mgh$ with $>40$ stars after implementing quality cuts discussed in Section~\ref{sec:methods}. Zero-point shifts have been included.}
\end{deluxetable*}

\end{document}